\documentclass[%
 reprint,
 amsmath,amssymb,
 aps,
]{revtex4-2}

\usepackage{graphicx}
\usepackage{dcolumn}
\usepackage{bm}
\usepackage{multirow}

\usepackage{hyperref}

\usepackage{csquotes}
\usepackage{amsmath}
\usepackage{mathtools, cuted}

\hypersetup{
    colorlinks,
    linkcolor={blue!80!black},
    citecolor={blue!80!black},
    urlcolor={blue!50!black}
}
\usepackage[utf8]{inputenc} 
\usepackage[mathlines]{lineno}

\usepackage{graphicx}
\usepackage{subcaption}
\usepackage{dcolumn}
\usepackage{tabularx}
\usepackage{multirow}
\usepackage{bm}
\usepackage{amsfonts,color} 
\usepackage[table,xcdraw]{xcolor}
\usepackage[english]{babel}
\usepackage{lipsum} 
\usepackage{array}
\usepackage{float}
\usepackage{adjustbox}
\usepackage{physics}

\definecolor{Gray}{gray}{0.9}

\newcolumntype{Y}{>{\centering\arraybackslash}X}
\newcolumntype{Z}{>{\raggedright\arraybackslash}X}
\newcolumntype{Q}{>{\raggedleft\arraybackslash}X}

\newcolumntype{P}[1]{>{\raggedright\arraybackslash}p{#1}}

\newcommand{\PreserveBackslash}[1]{\let\temp=\\#1\let\\=\temp}
\newcolumntype{C}[1]{>{\PreserveBackslash\centering}p{#1}}
\newcolumntype{L}[1]{>{\PreserveBackslash\raggedright}p{#1}}

\begin{document}

\preprint{APS/123-QED}

\title{Psychometric Evaluation of the Culture around Systemic Change Survey: A tool for Assessing Departmental Culture in Physics}

\author{Diana Sachmpazidi}
\author{Mike Verostek}

\affiliation{
School of Physics and Astronomy, Rochester Institute of Technology, NY 14623, USA\\
}%

\author{Jayna Petrella}

\affiliation{
 Department of Physics, University of Maryland, College Park, Maryland 20742, USA \\
}%

\author{Siwoo (Randy) Lee}
\affiliation{
 Department of Physics and Astronomy, University of Rochester, Rochester, New York 14627, USA \\
}%

\author{Chandra Turpen}
\affiliation{
 Department of Physics, University of Maryland, College Park, Maryland 20742, USA \\
}%





\begin{abstract}

Physics programs are continually evolving to better support student learning and meet the diverse needs of their populations. Achieving many of these goals requires not only structural adjustments but also fundamental shifts in departmental culture. Recognizing this, disciplinary organizations in physics have placed systemic change and equity at the center of reform efforts, identifying them as essential pillars of meaningful and sustainable change. Yet, tools for assessing departmental culture around educational change remain limited. In this study, we introduce the \textit{Culture around Systemic Change Survey} (CSCS), a new instrument designed to measure faculty and staff perceptions of their department's ``current'' and ``ideal''states. Using responses from the ``current'' scale  only ($N=111$ participants across 33 departments), we conducted a psychometric evaluation of the CSCS. Exploratory factor analysis supported a five-factor structure, including Open-Mindedness (OM), Student Involvement (SI), Collective Interpretation of Evidence (CE), Sustainability (S), and Disruption of Systemic Injustices (DI). As survey development is an iterative process, future work will focus on refinement and confirmatory analysis. This work sets the foundation for conducting population studies that assess the state of progress of the physics community along an equitable and systemic culture to pursuing educational change.

\end{abstract}

\keywords{retention model, graduate, doctoral, department support, mixed methods}
\maketitle


\section{\label{sec:intro}Introduction}

As higher education responds to shifting social, demographic, and institutional landscapes, academic departments, particularly in Science, Technology, Engineering, and Mathematics (STEM),  are being increasingly called upon to engage in not just responsive but transformative change \cite{census,kezar02}. In physics and related disciplines, this includes efforts to foster more inclusive and equitable environments and to critically examine departmental cultures that often reinforce the status quo.  National initiatives such as the  American Physical Society (APS) Inclusion Diversity and Equity Alliance (IDEA) \cite{aps_idea}, Effective Practices for Physics Programs (EP3) Departmental Action Leadership Institute (DALI) \cite{ep3_dali}, and American Institute of Physics (AIP) Task Force to Elevate African American Representation in Undergraduate Physics and Astronomy (TEAM-UP) \cite{aip_teamup} have placed systemic change and equity at the center of disciplinary advancement  \cite{foote14,hod17}. Yet, translating these goals into meaningful, sustained departmental action remains a significant challenge. 

Cultural change within departments requires more than adopting new programs or policies; it involves shifts in collective beliefs, values, and behaviors \cite{Schein04}. These deeper shifts, often called “second-order change" \cite{quan2019,Fry2014}, are essential to achieving goals that alter the culture of departmental life. Research on organizational change has shown that  successful efforts often depend on the ability of individuals and groups within a department to work collaboratively \cite{Kezar14}, reflect critically \cite{kezar02}, and adapt over time \cite{weick1999organizational}. 
However, departments often struggle to assess their current culture, monitor progress, or identify barriers to change \cite{eckel2003,Kezar14,sachmpazidi2024recognizing}.  

To address the need for better insight into departmental cultures, we developed the \textit{Culture around Systemic Change Survey (CSCS)}, an instrument designed to support national change efforts and help researchers understand the extent to which departmental actors pursue educational change through an equitable, systemic approach. Rather than relying solely on researcher-driven models, our development process was grounded in collaborative design. We partnered with, among others, leaders from multiple national initiatives to identify the cultural dimensions that stakeholders viewed as most critical. This collaborative approach ensured that the survey would be both relevant to on-going change work and practically useful to those supporting departments at scale.  

This project involves the development of two instruments: the current CSCS survey and the \textit{Culture around Equity and Inclusion Survey}, which will be the focus of a future publication. We view departmental cultures surrounding systemic change and those centered on equity and inclusion as deeply interconnected. As emphasized by APS-IDEA, advancing justice, equity, diversity, and inclusion depends on the collective capacity of departmental teams to transform their culture. A department's ability to respond to formative feedback around inclusion is closely tied to its readiness for systemic change. The CSCS was developed as a resource for national change initiatives to better understand patterns across departments, monitor progress, and provide targeted support based on each department's context. In this paper, we present the development and psychometric evaluation of the CSCS, conducted in summer 2024.

\section{\label{sec:background}Background}

Systemic change in higher education refers to a comprehensive and long-term process that fundamentally alters the typical operational framework of an organization. Rather than focusing on isolated fixes, systemic change spans multiple organizational domains and involves deep, pervasive transformation requiring shifts in culture \cite{quan2019,aaas2011,Fry2014,kezar02}. Because it addresses the underlying assumptions, norms, and power dynamics that shape institutional life, systemic change is often referred to as a \textit{second-order change}, distinguishing it from the more superficial or incremental \textit{first-order change} \cite{eckel2003}. 

Building on earlier distinctions between first- and second-order change, \citeauthor{reigeluth1994systemic} offer a helpful framing by contrasting \textit{piecemeal} and \textit{systemic change}. Piecemeal (or first-order) change modifies individual parts of a system without challenging its underlying structures or assumptions. In contrast, systemic (or second-order) change involves rethinking and redesigning the system as a whole to meet new goals and conditions. For example, consider a physics department aiming to improve graduate student retention. A first-order change might involve offering additional mentoring or tutoring sessions - practices that, while beneficial, leave core departmental structures and norms intact. In contrast, a second-order change could involve redesigning graduate-level core courses to incorporate more inclusive pedagogies, reexamining the department's advising model, or revising criteria for faculty promotion to reward inclusive mentorship practices. These changes require faculty to reflect critically on long-standing assumptions about teaching and departmental values.

The impact of such systemic efforts has been demonstrated in real-world contexts. One notable example is \textit{Strive Together}, a regional initiative based in Greater Cincinnati and Northern Kentucky, which brought together education, nonprofit, and civic leaders around a shared vision to improve student outcomes \cite{kania2011collective}. By fostering cross-sector collaboration and building a culture of shared accountability, Strive reported improvements in early childhood readiness, fourth-grade reading and math scores, and high school graduate rates. This example illustrates how coordinated, community-wide action grounded in systemic principles can lead to measurable and sustainable improvements on educational equity and success.

Systemic change is difficult precisely because it pushes against longstanding beliefs and norms that go unquestioned. It requires not just doing things differently, but thinking differently while critically reflecting on how departmental life should look like. As such, systemic change depends on a range of interrelated elements, including shared leadership, organizational learning, and the ability to interpret and respond to feedback. The following paragraphs expand on these elements that comprise a systemic approach to change. 

A systemic approach to change involves a diverse collective of individuals, including faculty, staff, and students, working together under a shared vision to achieve a common goal \cite{henderson2011facilitating}. Rather than rushing to implement solutions, these groups take a deliberate, reflective approach by first examining the root causes of the problems they aim to address \cite{kezar2013,ep3_dali,hora2017data}. This process typically involves gathering input form relevant stakeholders, such as students, and engaging in collective interpretation of the evidence to reflect transparent decision-making processes \cite{datnow2011collaboration}. 

Systemic change efforts also involve examining the institutions structures such as reward systems, leadership hierarchies that may reproduce inequities \cite{kezar02}. Throughout this process, they maintain regular communication with the broader department to foster transparency \cite{kezar2013}. Importantly, engaging in this type of change often requires difficult conversations about bias, privilege, and structural barriers that hinder equity and inclusion \cite{bensimon2005closing,bertschinger2020systemic,kezar2013}. Systemic change efforts unfold over iterative cycles of planning action, and reflection, reinforcing the idea that systemic change in not a one-time intervention but a long-term, adaptive process. By embracing these practices, departments build the capacity for deep, cultural change that characterizes systemic change \cite{Kezar14}. In this paper, we describe the development of a survey instrument designed to assess the extent to which department members perceive their department as engaging in key practices related to systemic approach for educational change, as well as their beliefs about whether these practices should be part of an ideal department.

\section{\label{sec:methods}Methods}

\subsection{\label{subsec:Survey_dev} Survey Development}

Traditional survey development often relies on expert-driven methods, which can overlook the lived experiences and needs of those most affected by the outcomes. Without input from key stakeholders—such as faculty, students, or department leaders—surveys may feel disconnected from local realities, leading to low engagement and limited impact. This gap is especially problematic in systemic change efforts, where cultural context and organizational dynamics are critical.

Our goal in developing this survey was to create a practical and relevant tool for leaders of APS and  AIP change initiatives. Specifically, the instrument is intended to help identify the resources, supports, and departmental readiness necessary for change, thereby fostering a more meaningful engagement and facilitative effective, tailored partnerships.  

To achieve this, we used a Human-Centered Design (HCD) approach \cite{cooley2000human}, which originates from engineering and product development and emphasizes the creation of tools based on a deep understanding of user needs. Guided by HCD principles, we organized a series of collaborative design sessions with 15 stakeholders holding multiple roles, including leaders of APS and AIP change initiatives, APS site visit leaders, department chairs, members of the APS Committee on Minorities (CoM) and Committee on the Status of Women in Physics (CSWP), representatives from AIP Research (formerly the Statistical Research Center), faculty members, and student representatives.

We conducted four virtual co-design sessions over the span of one month, each lasting two hours and held via Zoom. We used slides to guide the activities, and participants contributed to a shared Google document where notes were recorded in real time. These sessions provided a space for participants to collectively identify critical indicators of departmental culture related to systemic change and equity and inclusion. A more detailed description of the co-design process is provided in a separate publication \cite{SachmpazidiTurpenPERC2025}.

Following the co-design sessions, we synthesized the findings and reviewed relevant literature to inform the development of the main constructs. Each construct was clearly defined, and we developed an initial pool of items designed to capture multiple dimensions of each construct. We started with eight constructs:  \textit{Centering students' voices, Partnering with Students,	Advancing Equity and Inclusion, Shared Leadership, Transparency and Accountability of Practices of Change Teams, Centering Data and Informed Decision-Making, Context-Dependence,	Sustainability,} and \textit{Changing Hearts and Minds}. The DELTA instrument and the EP3 Chairs Survey served as key references \cite{ngai2020developing,chasteen2020results}. The DELTA instrument, in particular, guided the design of survey items that capture respondents’ perceptions of both an ideal departmental culture and the current departmental culture \cite{ngai2020developing}. 

Survey items were developed based on the departmental practices and values identified in the co-design sessions and adapted from the DELTA and EP3 Chairs surveys. The initial version of the CSCS consisted of 50 items measured on a 7-point response scale of `Strongly disagree to `Strongly agree.'  Each item asks respondents to consider the item statement in relation to how well it characterizes the ``current state of your department'' \textit{and} how well it characterizes the ``ideal department.''  Each item includes a pair of responses that can be compared to better understand respondents' current perceptions of their departmental culture, as well as what they would consider to be ideal qualities of a program. The items were coded into Qualtrics \cite{umdsurvey_ui}. We then conducted ten think aloud interviews with faculty and students. Although we initially intended for the survey to be completed by all department members, these interviews revealed that students could meaningful engage with only a subset of the items.  As a result, we narrowed the intended audience to faculty and staff. Following best practices in survey methodology \cite{blair2013designing}, we revised item wording, removed double-barreled questions, and ensured that language was clear and interpretable for faculty and staff respondents. 

Following the interviews, a panel of five survey experts reviewed the items for content validity \cite{beck2001ensuring}. Based on their feedback, we refined the wording and eliminated redundant items. The revised survey was then piloted across 33 physics departments, a process further described in detail in the remainder of this paper.

\subsection{\label{subsec:pilot} Pilot test}

\subsubsection{Physics Departments}



Our sampling source was the American Institute of Physics (AIP) list of U.S. institutions offering at least a bachelor's degree in physics ($N = 734$). From this dataset, we excluded institutions that were actively engaged in formal change initiatives at the time of sampling (APS-IDEA, EP3 DALI, or AIP TEAM-UP), yielding 620 eligible institutions. We piloted the survey in institutions not currently participating in these initiatives because the following pilot round is intended to be administered within change initiative-active institutions.

From the eligible pool, we randomly selected 33 institutions (28 Ph.D.-granting and 5 Bachelor's-only) for pilot administration. Of these 20 were public and 13 private. 
For each institution, we compiled faculty and staff contact lists from publicly available university directories and distributed the pilot survey via email to all individuals for whom addresses available.

\subsubsection{Sample of Participants}

 After deleting 18 cases from the original dataset, we ended up with $N=111$ responses (see Section \ref{subsubsec:missing} for a detailed description of which cases were removed and how missing data was handled). The demographic data of the sample are given in Table \ref{tab:demographics}.

\begin{table}[t]
\renewcommand\tabularxcolumn[1]{m{#1}}
\centering
\def\arraystretch{1.1}
\caption{\label{tab:demographics} Demographic data associated with the $N=111$ responses analyzed.  All participants were faculty and staff at one of 33 randomly selected physics bachelor's degree granting institutions that had not previously participated in a formal change initiative (APS-IDEA, EP3 DALI, or AIP TEAM-UP).}
\begin{tabularx}{\columnwidth}{>{\hsize=1.8\hsize}X
                               > {\hsize=.6\hsize}Y
                               > {\hsize=.6\hsize}Y}
\hline \hline
\textbf{Gender} & \textbf{\textit{N}} & \textbf{\%}  \\ 
\hline 
Man & 71 & 64\% \\ 
Woman & 24 & 22\%  \\
Genderqueer & 1 & 1\%  \\
Did not answer & 15 & 14\%  \\
\hline

\textbf{Race or ethnicity} & \textbf{\textit{N}} & \textbf{\%}   \\ 
\hline
White & 72 & 65\%  \\ 
Asian or Asian American & 8 & 7\%  \\
Hispanic or Latino & 2 & 2\%  \\
Black or African American & 1 & 1\%  \\
Middle Eastern or North African & 1 & 1\%  \\
Irish & 1 & 1\%  \\
Multiple races & 6 & 5\%  \\
Did not answer & 20 & 18\%  \\
\hline 

\textbf{Sexuality} & \textbf{\textit{N}} & \textbf{\%}  \\ 
\hline
Straight or Heterosexual & 86 & 77\%  \\ 
Queer & 15 & 14\%  \\
Did not answer & 10 & 9\%  \\
\hline 

\textbf{Position} & \textbf{\textit{N}} & \textbf{\%}  \\ 
\hline
Postdoc & 9 & 8\%  \\ 
Staff or Administration & 18 & 16\%  \\
Non-tenured and non-TT faculty & 13 & 12\%  \\
Tenured and TT faculty & 66 & 59\%  \\
Other & 5 & 5\% \\
\hline \hline

\end{tabularx}
\end{table}

\subsection{\label{subsec:dataanalysis} Data Analysis}

\subsubsection{\label{subsubsec:missing}Sample and Missing Data}  

Received 129 total responses (probably say more about this/recruitment etc).  4 were discarded because respondents exited the survey having answered fewer than 10 of the 50 items.  1 was identified for deletion due to consecutive identical responses throughout the survey, a type of ``careless response'' pattern that indicates the respondent's answers do not accurately reflect their beliefs \cite{meade2012identifying}.  

In the 124 remaining cases, 10.7\% of the data was missing.  Approximately 90\% of the missing data came from 27 respondents.  Much of the missingness occurred in the last 20 questions of the survey, and the highest missing response rate occurred on the final item ($N=32$ missing responses, 25.6\%), suggesting that at least some respondents may have exited due to the length of the instrument \cite{meade2012identifying, berry1992mmpi, brower2018too}. Item response distributions for the 27 cases from whom most of the missing data derived were qualitatively similar to distributions of responses from participants who completed the survey in its entirety.  This suggests a degree of similarity in how the two groups interpreted the survey, and lends credence to the idea that non-response was at least somewhat attributable to survey fatigue.  

However, there was a slightly higher rate of non-response on questions related to the experiences marginalized people in the department.  Some respondents may therefore have chosen not to answer these questions purposefully, perhaps if they felt these items espoused values that misaligned with their own.  We also observed that the rate of non-response among postdocs, staff, and non-tenured faculty was disproportionately high compared to their representation among the overall survey population.  In particular, among 13 individuals who did not respond to any of the items starting with the question stem ``Typically, systematic evidence about students' experiences in our program(s)...'', 8 held one of these positions. Some non-response may therefore have occurred because these individuals did not have firsthand knowledge of how evidence in their departments is used and felt unsure how to respond.

In the analysis presented in Results section \ref{subsec:results}, we opted to remove those 13 cases from the data, bringing the final sample size to $N=111$.  This decision was motivated by several preliminary versions of the exploratory factor analysis that suggested one of the factors (``Collective Evidence'') would consist almost entirely of items sharing the ``Typically, systematic evidence about students' experiences in our program(s)...'' stem.  Hence, data about this factor was effectively unavailable for those 13 respondents.  To better understand the effect of this decision, we conducted several versions of the EFA in which we included these cases using different missing data handling procedures.  However, regardless of whether those cases were included in the analysis or not, the overall factor structure remained highly consistent.  Comparisons of EFA results under different conditions are available Table \ref{tab:EFAcomparison} of the Appendix.  

In the final sample of $N=111$ responses presented in the results, 6.0\% of the data was missing.  Several common methods exist for handling missing data.  These often include listwise deletion, pairwise-deletion, and mean imputation.  However, such methods require a missing completely at random (MCAR) mechanism of generating the missing data, meaning that the probability of missing data has no relationship with any other values in the dataset \cite{bhaskaran2014difference}.  In our data, Little's MCAR test was statistically significant, indicating that the missingness is not MCAR \cite{little1988test} In fact, this is rarely a tenable assumption in real-world data \cite{rubin1976inference, muthen1987structural, nissen2019missing}. Even under the most ideal circumstances of MCAR data, evidence suggests that results may still be biased when using deletion methods or mean imputation \cite{brown1994efficacy, mazza2015addressing}.  

More typically, missing data are missing at random (MAR) or missing not at random (MNAR).  The critical difference between the two is that for MAR data, the probability of an observation being missing depends on other data in the dataset, but not on the missing data themselves \cite{bhaskaran2014difference}.  Here, we assume our data is MAR.  The higher rate of missingness among postdocs, staff, and non-tenured faculty indicated that at least some missingness was explained by other parts of our data.  Moreover, this type of missingness is amenable to the use of statistical methods for handling missing data, such as multiple imputation (MI) and full-information maximum likelihood (FIML) \cite{graham2015missing, van_buuren_flexible_2018, enders2001relative}.  Using such techniques allowed us to increase the statistical power of the study by avoiding further reduction of the sample size. 


We tested both MI and FIML methods for handling the remaining missing data and found results to be similar regardless of which was used (see Table \ref{tab:EFAcomparison} in the Appendix for a comparison of several methods discussed here).  One approach using MI was to generate a pooled covariance matrix, which was then used as the input for EFA in the \textit{psych} package in R \cite{nassiri2018using, psych_package, baseR_package}.  Another MI approach involved conducting a principal component analysis on each individual imputed dataset, and use generalized procrustes analysis to rotate the solutions toward a reference solution \cite{van2014using}.  Several FIML approaches were also tested, including use of FIML to generate an inter-item covariance matrix to use as input for EFA in the \textit{psych} package.  In the end, we opted to present the results of the EFA as carried out using FIML methods implemented in the \textit{umx} package \cite{umx_package}.  Since the results of each were mostly consistent and all were theoretically coherent, we decided to use the \textit{umx} package due to its ability to calculate individual factor scores for participants with missing responses.

\subsubsection{\label{subsubsec:assumptions_methods}Assumptions for EFA}

One significant consideration in the application of factor analysis is sample size, as correlation coefficients are likely to fluctuate more from sample to sample in smaller samples than large.  Hence, large sample sizes are generally desirable for more reliable factor analysis results, particularly if loadings on the factors are expected to be small \cite{knekta2019one}.  Common guidelines recommend a sample about 5 times the size of the number of items in the EFA, and more for missing data \cite{suhr2006exploratory, yong2013beginner, kass1979factor}.  For this survey, that would imply a sample size of approximately 250, just over twice the size of our sample ($N=111$).  However, other studies illustrate that this rule of thumb becomes less important as the magnitude of factor loadings increases, so long as factors are represented by enough variables to be clearly interpreted  \cite{arrindell1985empirical, guadagnoli1988relation}.  Indeed, \citeauthor{maccallum1999sample} showed that samples between 100 and 200 can be adequate when all communalities are greater than 0.5 (the communality of a variable is the part of its variance that can be accounted for by the common factors) \cite{maccallum1999sample}. Hence, potential disadvantages related to sample size are able to be mitigated through choice of retained factors, the cutoff value for factor loadings, and examination of communalities.

From a conceptual perspective, one critical assumption of EFA is that there is an underlying structure in the set of variables included in the analysis \cite{hair2010multivariate}.  In this regard, the extensive efforts described in Section \ref{subsec:Survey_dev} to develop the survey instrument support theoretical coherence between sets of items.  

From a quantitative perspective, one source of evidence for factorability is sufficiently large correlations between variables.  For preliminary analyses aimed at determining whether the data were suitable for EFA we leveraged the \textit{psych} package in R to generate a FIML inter-item correlation matrix.  Since survey responses were gathered on a seven-point Likert scale, we treated these variables as continuous in our analyses \cite{johnson1983ordinal, norman2010likert}.  Alongside examination of the correlation matrix, several metrics are typically recommended to judge the factorability of the dataset \cite{hair2010multivariate, jolliffe2002principal, knekta2019one, suhr2006exploratory, yong2013beginner}.  The Kaiser-Meyer-Olkin (KMO) measure of sampling adequacy represents the ratio of squared correlations between variables to squared partial correlation between variables, and is therefore an indicator of compact patterns of correlations \cite{kaiser1974little}.  Another common metric indicating suitability for factor analysis is Bartlett's test of sphericity, which tests whether the correlation matrix is significantly different from the identity matrix \cite{bartlett1937properties}.       

Although many results of factor analysis do not directly depend on specific distributional assumptions, departures from univariate and multivariate normality, linearity, and homoscedasticity are important due to their potential influence on the observed correlations between variables \cite{jolliffe2002principal, reise2000factor, hair2010multivariate}.  Mardia's multivariate normality test is one means of testing for violations of multivariate normality \cite{mardia1970measures}.  However, this test is of limited usefulness on its own, as even slight departures from non-normality in large samples can return a significant result \cite{kline2016principles}.  Thus, we also conducted a visual inspection of the data and calculated the skewness and kurtosis for each item's distribution.

Analyses testing the assumptions of EFA are presented in Results Section \ref{subsec:assumptions_results} and resulted in 4 items being removed from the EFA.  We therefore moved on to exploratory factor analysis using $N=111$ responses to the subset of 46 items deemed suitable for analysis.

\subsubsection{\label{subsubsec:efa_steps}Steps in EFA}
To choose the number of factors to retain in the exploratory factor analysis we relied on both theoretical considerations and several analytical indicators such as a scree plot (see Figure \ref{fig:screeplot}), Horn's parallel analysis, and Kaiser's criterion \cite{cattell1966scree, horn1965rationale, kaiser1960application}.  Each of these offer different means of determining the number of factors to retain.  Scree plots show the eigenvalues produced by a common factor model, and the number of factors to retain is indicated by an inflection point in the plot.  In parallel analysis, eigenvalues generated from the data are compared to the eigenvalues of randomly generated data with similar structure.  Components with larger eigenvalues than those of the randomly generated data are recommended to be retained  \cite{horn1965rationale}.  Meanwhile, Kaiser's criterion suggests retaining factors with eigenvalues greater than 1 \cite{kaiser1960application}.  

To support our decision regarding the number of factors to retain, we conducted several EFA's using different factor structures and examined their pattern matrices to determine which best supported our goal of a parsimonious and theoretically convergent factor structure.  Guidelines for practical and statistical significance for our sample size of $N=111$ suggest that factor loadings greater than approximately 0.5 are salient for interpretation \cite{norman2014biostatistics}.  However, this value is also dependent on the number of variables being analyzed, and smaller loadings are needed to be considered significant as the number of variables increases \citeauthor{hair2010multivariate}.  Furthermore, since this survey is intended for future use, a major goal of this exploratory factor analysis was to identify items for removal in future administrations. Given this goal, we were conscientious of retaining potentially meaningful items rather than strictly adhering to a cutoff point of 0.5. This approach allowed for the possibility that some items, which may not have loaded strongly in the current analysis due to the relatively small sample size, could become more significant with more data.  

Weighing these considerations, we chose to identify loadings of 0.40 or greater as significant.  Items that did not significantly load on any factor were systematically removed from the analysis until all remaining items were associated with a factor.  To be considered an adequate solution overall, we also required all factors to have a minimum of three items with significant loadings.  We applied these standards to the candidate solutions in order to arrive at the final five-factor solution detailed in Results Section \ref{subsubsec:efa} and summarized in Table \ref{tab:EFAresults}.  All analyses were carried out using FIML methods as implemented in the \textit{umx} package in R \cite{umx_package, baseR_package}.

\subsubsection{\label{subsubsec:reliability_methods}Reliability}
Reliability refers to the consistency of scores across multiple measurements of a variable \cite{aera2014standards, knekta2019one}.  However, when only one administration of an instrument is done, internal measures of consistency are used to estimate reliability \cite{aera2014standards}.  Most commonly, Cronbach's alpha is reported as an estimate of reliability and is given by 
\begin{equation}
    \alpha = \frac{n^2\overline{\sigma}_{ij}}{\sigma^2_x},
\end{equation}
where $\overline{\sigma}_{ij}$ is the average inter-item covariance, $\sigma^2_x$ is the variance of the scale score, and $n$ is the number of items in the scale \cite{cronbach1951coefficient, geldhof2014reliability}.  However, $\alpha$ relies on several assumptions that are often unrealistic in practice, including the assumption that all factor scores are the same for each item \cite{mcneish2018thanks}.  Therefore, in addition to reporting $\alpha$ as a measure of reliability, we also report we also report McDonald's $\omega_t$ and $\omega_h$ \cite{revelle2019reliability, mcdonald1999test}.  These model-based alternatives allow for greater insight into the internal structure of the instrument.  A brief discussion regarding the meaning and calculation of these $\omega$ coefficients using the \textit{psych} package in R is given here; however, for a more thorough theoretical description see Refs. \cite{gorsuch1983factor, mcdonald1999test, revelle2019reliability}.  For more in-depth guidance on calculating $\omega$ using R, see Refs. \cite{revelle2019reliability, rodriguez2016evaluating, flora2020your}).

\begin{figure*}[t]
\centering
\includegraphics[]{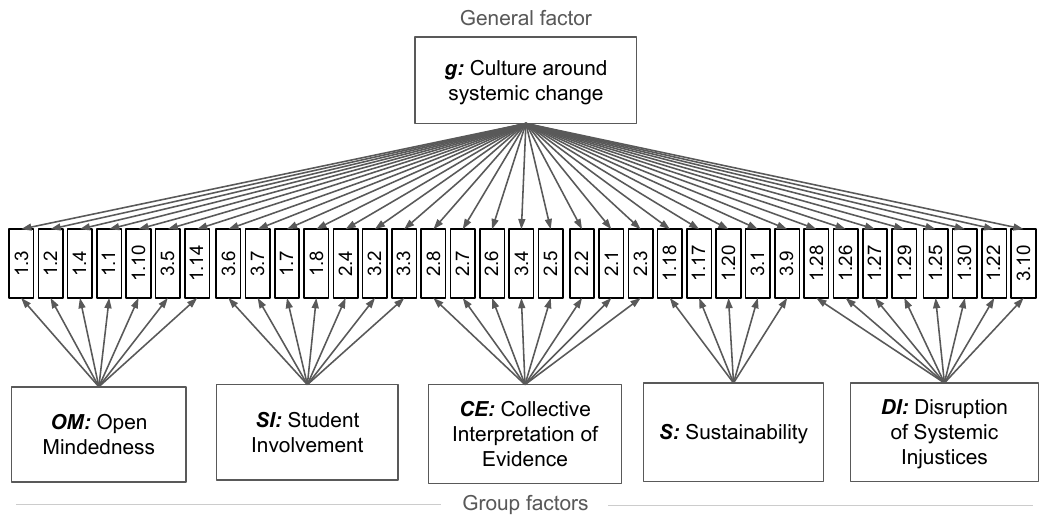}
\caption{\label{fig:general_factor} A graphical representation of the hierarchical model used to calculate $\omega_t$ and $\omega_h$.  Arrows pointing to items that did not significantly load on specific factors following Schmid-Leiman transformation are suppressed, although they may be nonzero.  Other means of estimating the general factor such as bifactor models force these loadings to zero (see \cite{reise2012rediscovery, rodriguez2016applying}).}
\end{figure*}

Factor analysis attempts to model the correlations between observed variables using a reduced set of latent unobserved variables (factors).  When an oblique rotation is used and the factors correlate with one another, a correlation matrix summarizing these relationships is produced (see Table \ref{tab:factor_cors}).  The factor correlation matrix can then be factored as well to find the loadings of these ``first order'' factors on a ``second order'' factor.  The first and second order factors can then be orthogonalized using the Schmid-Leiman transformation \cite{schmid1957development} to produce independent loadings on each observed variable due to both factor levels (see Figure \ref{fig:general_factor}). Conceptually, the second-order ``general'' factor measures what all the other first-order factors have in common.  In the context of this study, we interpret the general factor as measuring ``Culture around systemic change'' and representing what the group factors Open-mindedness (OM), Student Involvement (SI), Collective Interpretation of Evidence (CE), Sustainability (S), and Disruption of Systemic Inequities (DI) have in common. 

This hierarchical model serves to decompose the variance in the data into three sources: variance that is common to all items (general factor \textbf{g}), variance that is shared between specific groups of items (the set of orthogonal group factors \textbf{f}), and variance that is unique to each individual items (combined into a single error term \textbf{e}) \cite{revelle2019reliability, mcdonald1999test}.  Mathematically, if the observed data are given by a vector \textbf{x}, then
\begin{equation}
\textbf{x}=\textbf{cg}+\textbf{Af}+\textbf{e},
\end{equation}
where \textbf{c} is a column vector representing the factor loadings of each survey item on the general factor and \textbf{A} represents a matrix whose columns are the loadings of each survey item on a particular group factor.  

The reliability measures $\omega_t$ and $\omega_h$ are based on how the variance in the data is distributed across the general and group factors; that information is embedded in the loadings on those factors represented by \textbf{c} and \textbf{A} \cite{revelle2019reliability}.  $\omega_t$ is a model-based estimate of the proportion of variance attributable to all modeled sources of common variance, general and group \cite{rodriguez2016evaluating, revelle2019reliability}.  Analogous to coefficient $\alpha$, it is the proportion of true score variance to total variance \cite{mcdonald1999test, rodriguez2016applying} and is given by
\begin{equation}
\omega_t = \frac{\vb{1}'\vb{c}\vb{c}'\vb{1} + \vb{1}'\vb{A}\vb{A}'\vb{1}}{V_x},
\end{equation}
where $\vb{1}$ is a column vector of 1s, $\vb{1}'$ is its transpose, and $V_x$ is the total variance (the sum of all variances and covariances between items).  Equivalently, in terms of the  factor loadings,
\begin{equation} \label{eq:omegat2}
    \omega_t
         = \frac{
                \begin{aligned}
                    (\textstyle \sum\lambda_{gen})^2 + (\textstyle \sum\lambda_{grp1})^2 &+ (\textstyle \sum\lambda_{grp2})^2+...
                \end{aligned}
            }{V_x}
\end{equation}
and the total variance can be estimated as
\begin{align} \label{eq:varx}
\begin{split}
V_x = (\textstyle \sum\lambda_{gen})^2 &+ (\textstyle \sum\lambda_{grp1})^2 +\\ 
(\textstyle \sum\lambda_{grp2})^2&+...+ \textstyle \sum (1-h_i^2),
\end{split} 
\end{align}
where $\lambda_{gen}$, $\lambda_{grp1}$, $\lambda_{grp2}$... represent the factor loadings for the general and group factors and $h_i^2$ represents the communality of item $i$.  Together, Eqs. \ref{eq:omegat2} and \ref{eq:varx} illustrate that $\omega_t$ may be interpreted as a proportion of modeled variance to total variance.

Meanwhile, $\omega_h$ is a measure of the total score variance that is attributable only to the general factor \cite{rodriguez2016evaluating, revelle2019reliability}, and is therefore given by
\begin{equation} \label{eq:omegah}
\omega_h = \frac{\vb{1}'\vb{c}\vb{c}'\vb{1}}{V_x} = \frac{(\textstyle \sum\lambda_{gen})^2}{V_x}.
\end{equation}
The larger $\omega_h$, the more strongly the scale scores are influenced by the general factor common to all the indicators \cite{zinbarg2006estimating}.  Comparison of $\omega_t$ and $\omega_h$ therefore yield insight into the extent to which the reliable variance in total scores can be attributed to the general factor versus the group factors.  

Both $\omega_t$ and $\omega_h$ have analogue metrics, $\omega_{t \cdot sub}$ and $\omega_{h \cdot sub}$, for individual subscales. To refer to a specific subscale, we use its name in the metric's subscript (e.g., $\omega_{t \cdot grp1}$ refers to $\omega_{t \cdot sub}$ calculated for the group 1 subscale).  $\omega_{t \cdot sub}$ may be interpreted as the proportion of true score variance to total variance in a particular subscale.  For example, the two factors contributing to the modeled variance in the $grp1$ subscale are the general factor and the $grp1$ group factor.  Thus, 
\begin{equation} \label{eq:omegasub}
\omega_{t \cdot grp1} = \frac{(\textstyle \sum\lambda_{gen})^2 + (\textstyle \sum\lambda_{grp1})^2}{(\textstyle \sum\lambda_{gen})^2 + (\textstyle \sum\lambda_{grp1})^2 + \textstyle \sum (1-h_i^2)}.
\end{equation}
However, since Eq. \ref{eq:omegasub} reflects the proportion of variance attributable to both the general factor \textit{and} group factor, it may indicate high reliability even if the majority of that reliable variance stems from the general factor rather than from the group factor that the subscale is intended to measure. Hence, the subscale may seem to be a reliable indicator of a specific construct, when in fact its reliability largely reflects variance due to the general factor.

To address this issue we also compute $\omega_{h \cdot sub}$, which estimates the proportion of reliable variance in the subscale that can be attributed uniquely to the group factor after partitioning out the influence of the general factor.  Applying the logic of $\omega_h$ to Eq. \ref{eq:omegasub}, $\omega_{h \cdot grp1}$ is given by
\begin{equation} \label{eq:omegahsub}
\omega_{t \cdot grp1} = \frac{(\textstyle \sum\lambda_{grp1})^2}{(\textstyle \sum\lambda_{gen})^2 + (\textstyle \sum\lambda_{grp1})^2 + \textstyle \sum (1-h_i^2)}.
\end{equation}
When $\omega_{h \cdot sub}$ is low relative to $\omega_{t \cdot sub}$, much reliable variance in the subscale scores is attributable to the general factor rather than what is unique to the group factors \cite{rodriguez2016applying}.  This comparison provides a more accurate assessment of the extent to which the subscale score reflects variance specific to the intended construct, rather than variance shared across all items \cite{reise2013scoring}.  Calculation of \textbf{c} and \textbf{A} and their subsequent reliability estimates were carried out in R using the omega function in the \textit{psych} package.

\subsubsection{\label{subsubsec:validity_methods}Convergent and Discriminant Validity}
Validity refers to the extent to which an instrument measures the attribute of respondents that it is intended to measure \cite{aera2014standards, mcdonald1999test}.  Convergent validity is a measure of internal consistency of indicators measuring the same construct, while discriminant validity refers to the degree to which measures of different constructs are unique \cite{campbell1959convergent, bagozzi1991assessing}.  Evidence of convergent and discriminant validity is provided through several sources.  For example, one important aspect of convergent validity is whether individual items are interpreted by respondents in the way intended by the researcher \cite{knekta2019one}.  Meanwhile, a factor solution with few cross-loadings and moderate or low inter-factor correlations can be interpreted as evidence of discriminant validity.  

In addition to assessing these criteria, we also analyze several metrics of convergent and discriminant validity using the results of an independent clusters factor model fitted to the data.  Whereas EFA allows items to load onto more than one construct, an independent clusters model constrains observed variables to only load onto one factor \cite{mcdonald1999test} (results are shown in Table \ref{tab:indcluster}).  This model was fit to the data using the \textit{lavaan} package in R \cite{rosseel2012lavaan}.  Using these results, we employed the \citeauthor{fornell1981evaluating} criterion for evaluating convergent and discriminant validity.  This involves calculating the average variance extracted (AVE) for each construct, which measures the average amount of variance that a construct explains relative to the overall variance of its indicators.  AVE is calculated by averaging the squared factor loadings of the independent clusters model.  AVE values greater than 0.5 are indicative of convergent validity \cite{fornell1981evaluating}.

We then constructed an inter-factor correlation matrix by calculating the Pearson correlation between respondents' average construct scores.  We assessed discriminant validity by comparing the squared inter-construct correlations to the calculated AVE values.  Specifically, \citeauthor{fornell1981evaluating} suggest that for two constructs, discriminant validity is established when AVEs associated with both constructs are greater than their squared correlations.  This condition indicates that the latent variable explains more variance of the indicators than the other latent variable \cite{cheung2024reporting}.

\section{\label{subsec:results}Results}

\subsection{\label{subsec:assumptions_results} Assumptions for EFA}
Following the steps outlined in Methods Section \ref{subsubsec:assumptions_methods}, to test the suitability of the data for exploratory factor analysis we began by calculating Pearson correlations between the items.  Questions 1.5, 1.13, and 1.15 were observed to have correlations less than $r=0.3$ with over half of the other items, which raised concerns about their suitability for inclusion in a factor analysis.  Meanwhile, questions 1.23 and 1.24 had an inter-item correlation of 0.88, which suggested possible multicollinearity concerns for these items.

The overall KMO test value for our data was 0.89, supporting their general suitability for factor analysis.  Scores are on a 0 to 1 scale, with scores below 0.5 indicating the data are unfit for factor analysis and above 0.9 to be the `marvelous' \cite{kaiser1974little}.  Bartlett's test was significant at an alpha level of 0.05 ($\chi^2(1128)=5132.2$, $p<.001$), which further indicated that correlations between items were sufficient for factor analysis \cite{bartlett1937properties}.  

However, KMO scores for individual variables reinforced that questions 1.5, 1.13, and 1.15 were candidates for deletion, with scores of 0.71, 0.54, and 0.69, respectively.  Combined with evidence from the correlation matrix, we opted to eliminate questions 1.5, 1.13, 1.15, and 1.23 from the data prior to conducting a factor analysis.

Regarding distributional assumptions, Mardia's test was statistically significant and therefore indicated multivariate non-normality.  However, as noted in the methods, even slight departures from non-normality can return a significant result \cite{kline2016principles}.  We therefore analyzed the univariate distributions of each item (see Fig. \ref{fig:distributions} in the Appendix), which revealed that most had a skewness and kurtosis less than $\lvert 1.0 \rvert$.  All were under $\lvert 2.0 \rvert$ except question 1.14, which had a skewness of -1.76 and kurtosis of 3.33. These values indicated that the data was not severely non-normal \cite{bandalos2018factor, fabrigar2012exploratory}, and that maximum likelihood estimation was likely appropriate \cite{curran1996robustness}.  Finally, visual inspection of bivariate plots between each variable did not suggest any strong non-linear relationships, nor did the residual plots.  Thus, we deemed the data sufficient for exploratory factor analysis.

\subsection{\label{subsubsec:efa} Exploratory Factor Analysis}

\begin{figure}[t]
\centering
\includegraphics[]{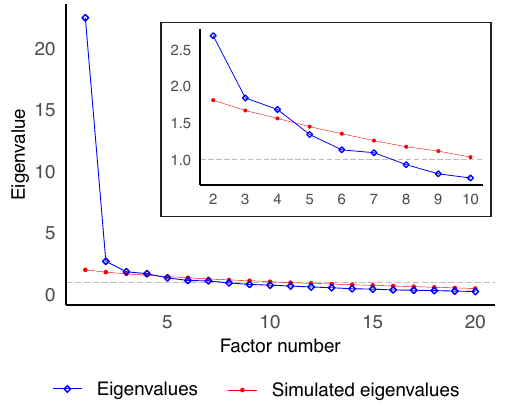}
\caption{\label{fig:screeplot} Scree plot showing eigenvalues produced by a common factor model.  The inset excludes the first eigenvalue to illustrate more detail.  The red line shows simulated eigenvalues used in parallel analysis, which suggested a 4-factor solution. The dotted grey line indicates eigenvalues above and below 1 (Kaiser's criterion).}
\end{figure}

The first step in exploratory factor analysis was to identify the appropriate number of factors to retain.  As discussed in Sections \ref{sec:background} and \ref{subsec:Survey_dev}, we initially theorized 8 underlying constructs. However, this analysis is exploratory, and we anticipated that several factors were likely to overlap conceptually, meaning that the number of distinct factors identified through EFA could be fewer.  

We started assessing the factor structure quantitatively by extracting eigenvalues produced by a common factor model and plotting them in the scree plot shown in Figure \ref{fig:screeplot}.  Parallel analysis was also conducted, in which the eigenvalues generated from the data are compared to the eigenvalues of randomly generated data with similar structure.  The eigenvalues from the randomly generated data are plotted in Figure \ref{fig:screeplot} as well.  The steep drop after the first eigenvalue might suggest that one factor could best fit the data.  However, this was not supported by theory, and other criteria indicated that more than one factor was appropriate.  Kaiser's criterion (retaining factors with eigenvalues greater than 1 \cite{kaiser1960application}) suggests a 7 factor solution, while parallel analysis indicates that 4 factors is the optimal choice.  These are highlighted in the inset plot of Figure \ref{fig:screeplot} by the red and gray lines.  Given several reasonable choices for factor retention, we ran models ranging from 4 to 8 factors, as well as a single factor model, to compare solutions for theoretical sense and total variance explained.  We also kept in mind the potential to overfit the data given the smaller sample size.  Due to the nature of the theorized constructs, we assumed that factors would be correlated and therefore used an oblique rotation (promax) to improve interpretation of solutions with multiple factors \cite{hendrickson1964promax}.

Next we used the criteria outlined in Methods Section \ref{subsubsec:efa_steps} (loadings of 0.40 or greater are significant, all factors must have a minimum of three significant loadings) to eliminate unsatisfactory solutions.  Solutions with 6 or more factors were quickly deemed inadequate due to too few significant loadings on several constructs, as well as concerns that these models would overfit the data.  This left the 4 and 5 factor solutions as candidates for selection.  However, since both potential solutions had a number of poorly fitting items, we continued our analysis by systematically removing those items with low pattern coefficients and re-running the analysis.  For both the 4 and 5 factor models, we removed items with low pattern coefficients ($<0.4$) in a stepwise process.  We began by eliminating items that did not seem to align well theoretically with its assigned factor, as well as those that we deemed were likely confusing or misinterpreted by respondents.  Cross-loaded items were considered individually.  Decisions to keep or remove these items were based on whether they clearly loaded more strongly on one factor than another.  In determining which items to remove, we also considered the item's communality.  Communalities lower than 0.50 were considered for deletion as recommended by \citeauthor{hair2010multivariate} \cite{hair2010multivariate}.  Choosing a solution for which each factor had most or all communalities greater than 0.5 was also important to ensure a stable solution given our sample size \cite{maccallum1999sample}.  Given the exploratory nature of this study and our aim to refine the survey instrument for future large-scale administration, we prioritized item retention at this stage. In close decisions about whether to retain an item, we erred on the side of inclusion, anticipating that future data collection will allow for more definitive decisions about item reduction.

Across both solutions, items 1.5, 1.6, 1.9, 1.11, 1.12, 1.13, 1.15, 1.16, 1.19, 1.21, 1.23, 1.24, and 3.8 were removed.  For the 4 factor model, questions 1.17 and 1.18 were also removed for failing to load onto any of the factors.  Comparing the solutions of the 4 and 5 factor models, we observed that three factors across both solutions were identical.  They represented three coherent themes: 1) collective decision-making by faculty and staff and their use of systematic evidence; 2) whether individuals with different backgrounds feel valued and respected in decision-making; and 3) open-mindedness to change.  The total variance explained was 59\% for the 4 factor model and 62\% for the 5 factor model.  From this perspective, it was not clear that adding a factor that only contributed 3\% to the explained variance was preferable to a more parsimonious model.  The decision to choose a 4 or 5 factor solution therefore depended on whether we believed the final factor in the 4 factor solution should be split into two.  

Theoretical considerations strongly influenced this decision-making process.  In the 4 factor model, the final factor included 10 items that we believed represented two separate constructs.  During item development, we had identified 3 of the items (1.20, 3.1, and 3.9) as having to do with sustainability of change; questions 1.17 and 1.18, which were dropped in the 4 factor model, were also associated with sustainability.  Meanwhile, the remaining 7 items were strongly associated with student involvement in decision-making processes.  A 5 factor solution served to split this final factor along this theoretical boundary, producing two factors with more coherent meaning.  Notably, items 1.17 and 1.18 were retained in the 5 factor model and belonged to the factor associated with sustainability of change as we originally theorized, providing evidence that 5 factors best fit the data.  Adding this factor also improved the communalities for its associated items.  For instance, the communalities for items 3.1 and 3.9 were raised from 0.44 and 0.38 in the 4 factor solution to 0.60 and 0.53 in the 5 factor solution. 

The final 5 factor solution is presented in Table \ref{tab:EFAresults}.  Once a satisfactory solution had been found, we finalized our interpretation of the 5 factors and selected names for each. They were labeled Open-mindedness (OM), Student Involvement (SI), Collective Interpretation of Evidence (CE), Sustainability (S), and Disruption of Systemic Injustices (DI).  Definitions are given in Table  \ref{tab:definitions}.  This solution accounted for 69\% of the total variance in the 35 items.  Specifically, OM ($n=7$) accounted for 11\%, SI ($n=7$) accounted for 19\%, CE ($n=8$) accounted for 14\%, S ($n=5$) accounted for 11\%, and DI ($n=8$) accounted for 14\%.  

\subsection{\label{subsubsec:example_analysis} Example Data Analysis}
In this section, we illustrate several examples of ways the data collected with this instrument might be used by departmental change leaders to inform decision-making.  For example, leaders may be interested in answering the question, ``How are departments’ current cultural practices around systemic change assessed by their faculty and staff members, and are there any areas that are seen as strengths?''  We used the data from this paper to explore this question in as part of a previous publication \cite{verostekPERC2025}, but we briefly outline the methods and results here.    

\begin{figure}[h]
\centering
\includegraphics[]{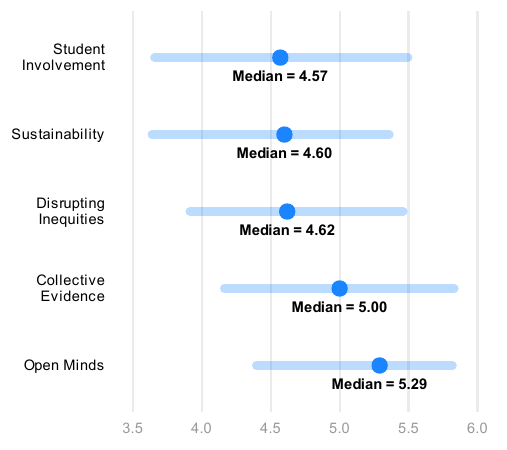}
\caption{\label{fig:current_iqr} Interquartile ranges of factor scores on the ``current'' scale.  Respondents score their departments significantly higher on the Open Minds and Collective Evidence factors than on Student Involvement, Sustainability, or Disrupting Inequities.}
\end{figure}

First we calculated respondent's composite factor scores based on the average of the items within each factor.  We then calculated the mean, standard deviation, and median values of these factor scores across all respondents.  These are summarized in Table \ref{tab:summary_stats}.  Distributions of factor scores are shown in Figure \ref{fig:current_iqr} and visually suggest that the Open Minds factor and Collective Evidence factor may have higher scores on average than the other three factors (Sustainability, Student Involvement, and Disrupting Injustices). We performed a one-way repeated-measures ANOVA to test whether a significant difference exists between mean current factor scores.  Results of the ANOVA were significant, $F(4,440)=13.90$, $p<0.001$.

Then, to explore differences between specific pairs of factors, we use paired $t$-tests with a Bonferroni corrected $p$-value ($p_{adj}$=$10p$) to test for significance at the $\alpha=0.05$ level.  We also report the point-biserial correlation $r$ as a measure of effect size, calculated as $\sqrt{t^2/(t^2+df)}$ where $t$ is the test statistic \cite{fritz2012effect}.  Those results reveal that on average, participants rate their current department significantly higher on the Open Minds factor ($M=4.97$, $SD=1.22$) than on Student Involvement ($M=4.45$, $SD=1.35$), $t(110)=5.44$, $p_{adj}<0.001$, $r=0.46$; Sustainability ($M=4.44$, $SD=1.32$), $t(110)=5.36$, $p_{adj}<0.001$, $r=0.45$; and Disrupting Injustices ($M=4.54$, $SD=1.30$), $t(110)=4.95$, $p_{adj}<0.001$, $r=0.43$.  These represent medium effect sizes according to guidelines set by Cohen ($r=0.10$ represents a small effect, $r=0.30$ a medium effect, and $r=0.50$ a large effect) \cite{cohen1988statistical, cohen1992power}.  Similar to the Open Minds factor, participants rate their current department significantly higher on the Collective Evidence factor ($M=4.85$, $SD=1.26$) than on Student Involvement, $t(110)=4.20$, $p_{adj}<0.001$, $r=0.37$; Sustainability, $t(110)=4.64$, $p_{adj}<0.001$, $r=0.40$; and Disrupting Injustices, $t(110)=3.58$, $p_{adj}=0.005$, $r=0.32$.  All of these represent medium effect sizes.  The average difference between scores on the Open Minds factor and the Collective Evidence factor was not statistically significant, nor were any of the pairwise comparisons between the Student Involvement, Sustainability, and Disrupting Injustices factors.  In summary, these results indicate that respondents score their departments significantly higher on the Open Minds and Collective Evidence factors than on Student Involvement, Sustainability, or Disrupting Inequities.

The relatively higher ratings for Open Minds and Collective Evidence may suggest that faculty and staff broadly view their departments as open to learning and willing to use data collaboratively in order to guide change.  For change leaders, together these results may be leveraged as foundational resources when planning how to best approach a new change initiative.  A departmental culture that is receptive to new ideas and willing to engage in data-based decision-making could provide a fertile environment for significant and sustained transformation, particularly when there is data to support a shared view that change is needed.

\begin{table}[t]
\renewcommand\tabularxcolumn[1]{m{#1}}
\def\arraystretch{1.1}
\begin{tabularx}{\columnwidth}{>{\hsize=.6\hsize}X
                               > {\hsize=1.1\hsize}Y
                               > {\hsize=1.1\hsize}Y
                               > {\hsize=1.1\hsize}Y
                               > {\hsize=1.1\hsize}Y}

\hline \hline
 & Mean & SD & Median & \% Var \\  
 \hline

\textbf{OM} & 4.97 & 1.22 & 5.29 & 11\%  \\ 
\textbf{SI} & 4.45 & 1.35 & 4.57 & 19\%  \\ 
\textbf{CE} & 4.85 & 1.26 & 5.00 & 14\%   \\ 
\textbf{S} & 4.44 & 1.32 & 4.60 & 11\%  \\ 
\textbf{DI} & 4.54 & 1.30 & 4.62 & 14\%   \\ 

\hline\hline
\end{tabularx}
\caption{\label{tab:summary_stats} Summary statistics describing the distribution of factor scores across the five constructs identified through EFA.  For this study, factor scores for individual participants are calculated as the mean score of the items associated with a factor.}
\end{table}

\begin{table}[t]
\renewcommand\tabularxcolumn[1]{m{#1}}
\centering
\def\arraystretch{1.1}
\begin{tabularx}{\columnwidth}{>{\hsize=1\hsize}X
                               > {\hsize=1\hsize}Y
                               > {\hsize=1\hsize}Y
                               > {\hsize=1\hsize}Y
                               > {\hsize=1\hsize}Y
                               > {\hsize=1\hsize}Y}
\cline{1-6}
 & \textbf{OM} & \textbf{SI} & \textbf{CE} & \textbf{S} & \textbf{DI}  \\ \hline 

\textbf{OM} & 1.00 &  &  &  &  \\ 
\textbf{SI} & 0.60 & 1.00 &  &  &  \\ 
\textbf{CE} & 0.55 & 0.72 & 1.00 &  &  \\ 
\textbf{S} & 0.52 & 0.74 & 0.73 & 1.00 &  \\ 
\textbf{DI} & 0.49 & 0.67 & 0.58 & 0.58 & 1.00 \\ 

\hline
\end{tabularx}
\caption{\label{tab:factor_cors} Factor correlations as determined by the oblique rotation in EFA, often denoted as $\phi$.  Oblique rotations allow the factors to correlate, meaning the new axes in the subspace of retained factors are not constrained to be perpendicular to one another.  The values in the correlation matrix are analogous to the dot product between these axes \cite{gorsuch1983factor}.}
\end{table}

\subsection{Validity and Reliability of Sample Data}

Regarding \citeauthor{fornell1981evaluating}'s metrics for convergent and discriminant validity, we find values of AVE for each construct to be greater than the threshold value of 0.50 to indicate convergent validity \cite{fornell1981evaluating}.  We also found that the squared inter-construct correlation values were all lower than the AVE values for each construct, providing evidence for discriminant validity.  These results are summarized in Table \ref{tab:validity}.     

We assessed internal consistency by calculating reliability coefficients $\alpha$, $\omega_t$, and $\omega_h$ for both the whole instrument as well as its individual subscales.  The results of these calculations are summarized in Table \ref{tab:reliability_stats}.  For the total score, we find $\omega_t=0.98$ and $\alpha=0.97$, which indicate nearly all of the observed score variance in the 35 items can be attributed to ``true score'' variance.  We also find $\omega_h=0.88$, which indicates that 88\% of the total score variance is attributable to a general factor.  This means that only about 10\% (0.98-0.88) of the reliable variance can be attributed to the multidimensionality of the group factors.  This indicates that the total score reflects the general factor well, and the total score can be regarded as a sufficiently reliable measure of the general factor \cite{reise2013scoring}.

We also calculated $\omega_{t \cdot sub}$, $\omega_{h \cdot sub}$, $\alpha_{sub}$ for the individual subscales to better understand the extent to which they are reliably measuring their specific intended construct.  We find $\omega_{t \cdot sub}$ and $\alpha_{sub}$ scores all between 0.89 and 0.95, indicating that the proportion of true score variance in the subscales is also quite high.  However, calculations of $\omega_{h \cdot sub}$ are much lower.  The highest omega hierarchical subscale is $\omega_{h \cdot OM}=0.40$ and the lowest are $\omega_{h \cdot SI}$ and $\omega_{h \cdot CE}$, which are equal to $0.18$.  These values represent the common variance remaining after partitioning out the contribution of the general factor.  They reflect the degree to which the subscale score reflects the intended specific factor.  \citeauthor{reise2013scoring} suggest $\omega_{h \cdot sub}$ should be greater than $0.50$ in order to reliably use the subscales as a measure of their intended construct \cite{reise2013scoring}.  Presently, taking the values of $\omega_{t \cdot sub}$ and $\omega_{h \cdot sub}$ together, we conclude that the high reliability of the subscale scores indicated by $\omega_{t \cdot sub}$ is mostly attributable to individual differences on the general factor rather than differences on the subscales.  Thus, further evidence of subscale reliability should be established before confidently drawing conclusions from individual subscale scores.

We anticipate that the subscale reliabilities will improve as we continue to refine this instrument in future iterations of data collection.  This is because, in line with the exploratory goals of the study, we retained some borderline items including several with cross-loadings. This decision reflected our intention to conduct a more comprehensive item evaluation in future analyses based on larger-scale data.  However, inclusion of such items here likely diminished subscale reliability \cite{reise2013scoring}.  Indeed, in a separate analysis of this data in which we moved the threshold for item inclusion to a loading of 0.50 and removed the cross-loaded items, all $\omega_{h \cdot sub}$ values aside from $\omega_{h \cdot SI}$ increased to between 0.40 and 0.60 without significant reduction in total reliability.

\begin{table}[t]
\renewcommand\tabularxcolumn[1]{m{#1}}
\def\arraystretch{1}
\begin{tabularx}{\columnwidth}{>{\hsize=.7\hsize}X
                               > {\hsize=1.1\hsize}Y
                               > {\hsize=1.1\hsize}Y
                               > {\hsize=1.1\hsize}Y}

\hline \hline

 & $\omega_t$ & $\omega_h$ & $\alpha$    \\
\textbf{Total} & 0.98 & 0.88 & 0.97    \\ 
\hline
 & $\omega_{t \cdot sub}$ & $\omega_{h \cdot sub}$ & $\alpha_{sub}$    \\
\textbf{OM} & 0.91 & 0.40 & 0.89    \\ 
\textbf{SI} & 0.93 & 0.18 &  0.93    \\ 
\textbf{CE} & 0.95 & 0.18 & 0.95   \\ 
\textbf{S} & 0.90 & 0.23 & 0.89   \\ 
\textbf{DI} & 0.94 & 0.30 & 0.93  \\ 

\hline\hline
\end{tabularx}
\caption{\label{tab:reliability_stats} Summary of reliability statistics across the entire instrument (top row) and individual subscales.  The high values of omega total and alpha indicate excellent internal consistency.  The comparatively lower values of $\omega_{h \cdot sub}$ mean that further evidence of subscale reliability should be established before confidently drawing conclusions from individual subscale scores.}
\end{table}

\begin{table}[t]
\renewcommand\tabularxcolumn[1]{m{#1}}
\centering
\def\arraystretch{1.1}
\begin{tabularx}{\columnwidth}{>{\hsize=1\hsize}X
                               > {\hsize=1\hsize}Y
                               > {\hsize=1\hsize}Y
                               > {\hsize=1\hsize}Y
                               > {\hsize=1\hsize}Y
                               > {\hsize=1\hsize}Y}
\cline{1-6}
 & \textbf{OM} & \textbf{SI} & \textbf{CE} & \textbf{S} & \textbf{DI}  \\ \hline 

\textbf{OM} & (0.57) &  &  &  &  \\ 
\textbf{SI} & 0.47 & (0.65) &  &  &  \\ 
\textbf{CE} & 0.45 & 0.57 & (0.70) &  &  \\ 
\textbf{S} & 0.33 & 0.44 & 0.57 & (0.64) &  \\ 
\textbf{DI} & 0.31 & 0.44 & 0.47 & 0.34 & (0.65) \\ 

\hline
\end{tabularx}
\caption{\label{tab:validity} Matrix summarizing \citeauthor{fornell1981evaluating}'s metrics for convergent and discriminant validity \cite{fornell1981evaluating}.  The diagonals represent the average variance extracted for each construct while off-diagonals represent the squared Pearson correlations between participants' scores on the different factors.  Squared inter-construct correlations were all lower than the AVE values, providing evidence for discriminant validity. Meanwhile, AVE values were above 0.50, providing evidence for convergent validity.}
\end{table}

\begin{table*}[t]
\renewcommand\tabularxcolumn[1]{m{#1}}
\centering
\def\arraystretch{1.0}
\caption{\label{tab:EFAresults} The final 5 factor solution as estimated using the umx package in R.  Promax rotation was used.  Bold indicates that the factor loading was above 0.40, which we determined as the cutoff for a significant factor loading.}
\begin{tabularx}{\textwidth}{>{\hsize=4.5\hsize}X
                               > {\hsize=.3\hsize}X
                               > {\hsize=.3\hsize}X
                               > {\hsize=.3\hsize}X
                               > {\hsize=.3\hsize}X
                               > {\hsize=.3\hsize}X}
\cline{1-6}
 &  \multicolumn{5}{c}{Loading} \\
\cline{2-6}

\multicolumn{1}{l}{Item number and content by factor} & \multicolumn{1}{c}{1} & \multicolumn{1}{c}{2} & \multicolumn{1}{c}{3} & \multicolumn{1}{c}{4} & \multicolumn{1}{c}{5} \\ \hline 

\multicolumn{1}{l}{Factor 1 --- Open to engaging with and learning from others [OM]} & & & & & \\ 
\textbf{On average, people in my department} & &  &  &  &  \\ 
\hspace{3mm}(1.3) are open to revising their thinking. & \hphantom{$-$}$\textbf{0.99}$ & $-0.15$ & $-0.08$ & \hphantom{$-$}$0.02$ & \hphantom{$-$}$0.09$ \\ 
\hspace{3mm}(1.2) take advantage of opportunities to learn, grow, or change. & \hphantom{$-$}$\textbf{0.79}$  & \hphantom{$-$}$0.29$  & $-0.09$  & \hphantom{$-$}$0.07$  & $-0.16$  \\ 
\hspace{3mm}(1.4) hold stubbornly to their own opinion. & \hphantom{$-$}$\textbf{0.77}$  & \hphantom{$-$}$0.02$ & $-0.05$ & $-0.25$ & \hphantom{$-$}$0.18$ \\ 
\hspace{3mm}(1.1) engage with differing perspectives. & \hphantom{$-$}$\textbf{0.69}$ & \hphantom{$-$}$0.17$ & $-0.11$ & \hphantom{$-$}$0.06$ & $-0.09$ \\ 
\hspace{3mm}(1.10) share the reasoning behind the changes made. & \hphantom{$-$}$\textbf{0.50}$  & $-0.01$  & $-0.17$  & \hphantom{$-$}$0.32$  & \hphantom{$-$}$0.04$  \\ 
\textbf{On average, my department’s change efforts}  &  &  &   &  &  \\
\hspace{3mm}(3.5) are driven by a shared responsibility among department faculty for the health of 

\hspace{3mm}the department and the people in it. & \hphantom{$-$}$\textbf{0.44}$\newline & $-0.02$\newline  & \hphantom{$-$}$0.39$\newline  & \hphantom{$-$}$0.03$\newline  & \hphantom{$-$}$0.10$\newline \\
\textbf{On average, people in my department}
 &   &  &  &   &   \\
\hspace{3mm}(1.14) take into account the current state of our program(s) when planning 

\hspace{3mm}a change effort.
 & \hphantom{$-$}$\textbf{0.42}$\newline  & \hphantom{$-$}$0.24$\newline  & \hphantom{$-$}$0.20$\newline  & \hphantom{$-$}$0.05$\newline  & $-0.05$\newline \\

\cline{1-6}
\multicolumn{1}{l}{Factor 2 --- Student involvement in change efforts [SI]} & & & & &  \\
\textbf{On average, my department’s change efforts} &  &  &  &  \\ 
\hspace{3mm}(3.6) involve students in decision making. & \hphantom{$-$}$0.05$ & \hphantom{$-$}$\textbf{0.95}$ & \hphantom{$-$}$0.04$ & $-0.26$ & \hphantom{$-$}$0.00$  \\ 
\hspace{3mm}(3.7) involve students in implementing changes. & \hphantom{$-$}$0.11$ & \hphantom{$-$}$\textbf{0.92}$ & \hphantom{$-$}$0.04$ & $-0.11$ & $-0.18$ \\
\textbf{On average, people in my department}  &  &  &  &  \\
\hspace{3mm}(1.7) partner with student representatives to collectively pursue change efforts.   & \hphantom{$-$}$0.02$ & \hphantom{$-$}$\textbf{0.91}$ & $-0.17$ & $-0.01$ & \hphantom{$-$}$0.09$ \\
\hspace{3mm}(1.8) partner with student representatives in a way that allows students to meaningfully 

\hspace{3mm}participate in decision making.  & $-0.10$\newline & \hphantom{$-$}$\textbf{0.84}$\newline & $-0.07$\newline & $-0.14$\newline & \hphantom{$-$}$0.19$\newline \\
\textbf{Typically, systematic evidence about students’ experiences in our program(s)}  & &  &  &  &\\
\hspace{3mm}(2.4) is interpreted in collaboration with student representatives.  & $-0.13$ & \hphantom{$-$}$\textbf{0.79}$ & \hphantom{$-$}$0.04$ & $-0.03$ & \hphantom{$-$}$0.11$ \\
\textbf{On average, my department’s change efforts}   &  &  &  &  &  \\
\hspace{3mm}(3.2) are based on adapting research-based, well-established practices to local situations.  & \hphantom{$-$}$0.14$ & \hphantom{$-$}$\textbf{0.65}$ & \hphantom{$-$}$0.03$ & \hphantom{$-$}$0.14$ & $-0.04$ \\
\hspace{3mm}(3.3) are informed by our students’ values or goals.  & \hphantom{$-$}$0.20$ & \hphantom{$-$}$\textbf{0.56}$ & \hphantom{$-$}$0.03$ & \hphantom{$-$}$0.22$ & $-0.06$ \\

\cline{1-6}
\multicolumn{1}{l}{Factor 3 --- Change through collective interpretation of evidence [CE]} & & & & & \\
\textbf{Typically, systematic evidence about students’ experiences in our program(s)}  &  &  &  &  &  \\ 
\hspace{3mm}(2.8) guides collective actions by faculty and/or staff.  & $-0.18$ & \hphantom{$-$}$0.02$ & \hphantom{$-$}$\textbf{0.95}$ & \hphantom{$-$}$0.02$ & \hphantom{$-$}$0.11$ \\ 
\hspace{3mm}(2.7) leads to collective decision making among faculty, and/or staff.  & $-0.09$ & \hphantom{$-$}$0.04$ & \hphantom{$-$}$\textbf{0.95}$ & $-0.04$ & \hphantom{$-$}$0.07$ \\
\hspace{3mm}(2.6) is interpreted in a collaborative environment with multiple faculty members.  & $-0.02$ & \hphantom{$-$}$0.03$ & \hphantom{$-$}$\textbf{0.77}$ & \hphantom{$-$}$0.00$ & \hphantom{$-$}$0.16$ \\
\textbf{On average, my department’s change efforts}  &  &  &  &  &  \\
\hspace{3mm}(3.4) are based on consensus among faculty.  & \hphantom{$-$}$0.39$ & $-0.36$ & \hphantom{$-$}$\textbf{0.74}$ & $-0.09$ & \hphantom{$-$}$0.05$ \\
\textbf{Typically, systematic evidence about students’ experiences in our program(s)}  &  &  &  &  \\
\hspace{3mm}(2.5) is used to identify the nature of problems in our program(s).  & $-0.04$ & \hphantom{$-$}$0.35$ & \hphantom{$-$}$\textbf{0.49}$ & \hphantom{$-$}$0.12$ & $-0.02$ \\
\hspace{3mm}(2.2) leads to change(s) (e.g. modifying recruitment practices, curricular modifications, 

\hspace{3mm}instructional changes, etc.).  & $-0.04$\newline & \hphantom{$-$}$0.37$\newline & \hphantom{$-$}$\textbf{0.47}$\newline & \hphantom{$-$}$0.27$\newline & $-0.23$\newline \\
\hspace{3mm}(2.1) leads to informed decision making.  & \hphantom{$-$}$0.09$ & \hphantom{$-$}$0.36$ & \hphantom{$-$}$\textbf{0.45}$ & \hphantom{$-$}$0.05$ & $-0.04$ \\
\hspace{3mm}(2.3) is used to improve the experiences of marginalized students.  & $-0.05$ & \hphantom{$-$}$0.30$ & \hphantom{$-$}$\textbf{0.40}$ & \hphantom{$-$}$0.03$ & \hphantom{$-$}$0.27$ \\

\cline{1-6}
\multicolumn{1}{l}{Factor 4 --- Systematic approaches to planning change and monitoring changes [S]} & & & & & \\
\textbf{On average, people in my department} &  &  &  &  &  \\ 
\hspace{3mm}(1.18) build assessments into change plans.  & $-0.03$ & $-0.25$ & $-0.01$ & \hphantom{$-$}$\textbf{0.98}$ & \hphantom{$-$}$0.12$ \\ 
\hspace{3mm}(1.17) systematically monitor change efforts to understand progress toward goals.  & \hphantom{$-$}$0.10$ & $-0.16$ & $-0.13$ & \hphantom{$-$}$\textbf{0.94}$ & \hphantom{$-$}$0.11$ \\
\hspace{3mm}(1.20) build plans for how to overcome potential challenges with change initiatives. & \hphantom{$-$}$0.08$ & \hphantom{$-$}$0.10$ & \hphantom{$-$}$0.07$ & \hphantom{$-$}$\textbf{0.70}$ & $-0.03$ \\
\textbf{On average, my department’s change efforts}  &  &  &  &  &  \\
\hspace{3mm}(3.1) are supported by formal evidence (such as outcomes from surveys, interviews, 

\hspace{3mm}literature, expert feedback, assessments, national or institutional reports).  & $-0.04$\newline & \hphantom{$-$}$0.24$\newline & \hphantom{$-$}$0.17$\newline & \hphantom{$-$}$\textbf{0.54}$\newline & $-0.13$\newline \\
\hspace{3mm}(3.9) are documented for others to consider or review.  & $-0.17$ & \hphantom{$-$}$0.21$ & \hphantom{$-$}$0.18$ & \hphantom{$-$}$\textbf{0.46}$ & \hphantom{$-$}$0.06$ \\

\cline{1-6}
\multicolumn{1}{l}{Factor 5 --- Fostering inclusive change to disrupt systemic injustices [DI]} & & & & & \\
\textbf{On average, people in my department}  &  &  &  &  &  \\ 
\hspace{3mm}(1.28) take steps to ensure that marginalized people feel safe and comfortable

\hspace{3mm}voicing their opinions or concerns.  & \hphantom{$-$}$0.08$\newline & $-0.18$\newline & \hphantom{$-$}$0.22$\newline & $-0.07$\newline & \hphantom{$-$}$\textbf{0.90}$\newline \\ 
\hspace{3mm}(1.26) take steps to ensure that marginalized people have an active role in 

\hspace{3mm}departmental decision making.
 & \hphantom{$-$}$0.04$\newline & $-0.09$\newline & \hphantom{$-$}$0.02$\newline & \hphantom{$-$}$0.06$\newline & \hphantom{$-$}$\textbf{0.85}$\newline \\
\hspace{3mm}(1.27) actively attend to the needs of marginalized people.  & $-0.03$ & \hphantom{$-$}$0.01$ & $-0.02$ & \hphantom{$-$}$0.15$ & \hphantom{$-$}$\textbf{0.83}$ \\
\hspace{3mm}(1.29) address power differentials in departmental conversations.  & $-0.02$ & \hphantom{$-$}$0.24$ & \hphantom{$-$}$0.12$ & $-0.11$ & \hphantom{$-$}$\textbf{0.62}$ \\
\hspace{3mm}(1.25) monitor the effects of the change efforts on the experiences of marginalized people. & $-0.09$ & \hphantom{$-$}$0.01$ & \hphantom{$-$}$0.03$ & \hphantom{$-$}$\textbf{0.40}$ & \hphantom{$-$}$\textbf{0.59}$\\
\hspace{3mm}(1.30) take action to build a more just system.  & \hphantom{$-$}$0.18$ & \hphantom{$-$}$0.33$ & $-0.12$ & $-0.03$ & \hphantom{$-$}$\textbf{0.55}$ \\
\hspace{3mm}(1.22) disrupt biases that they recognize in departmental processes. & \hphantom{$-$}$0.24$ & \hphantom{$-$}$0.19$ & $-0.10$ & \hphantom{$-$}$0.07$ & \hphantom{$-$}$\textbf{0.50}$ \\
\textbf{On average, my department’s change efforts} &  &  &  &  &  \\
\hspace{3mm}(3.10) are designed to rectify past injustices (e.g. discrimination, exclusion). & $-0.16$ & \hphantom{$-$}$0.34$ & $-0.02$ & \hphantom{$-$}$0.08$ & \hphantom{$-$}$\textbf{0.47}$ \\
\hline
\end{tabularx}
\end{table*}

\begin{table*}[t]
\renewcommand\tabularxcolumn[1]{m{#1}}
\centering
\def\arraystretch{1.0}
\begin{tabularx}{\textwidth}{>{\hsize=1.0\hsize}X}

\hline\hline
\textbf{1 --- Open to engaging with and learning from others [Open Minds (OM) 7 items]} \\ 
\textit{Definition:} Willingness of department members to consider new perspectives and revise their thinking in an effort to improve the overall well-being of the community (\textit{e.g.,} ``On average, people in my department take advantage of opportunities to learn, grow, or change.'')\\
\hline
\textbf{2 --- Student involvement in change efforts [Student Involvement (SI), 7 items]} \\ 
\textit{Definition:} Extent to which students are actively engaged in departmental change efforts through collaboration with faculty and staff and input in decision-making to ensure changes are informed by students' values and goals (\textit{e.g.,} ``On average, my department's change efforts involve students in decision making.'')\\
\hline
\textbf{3 --- Change through collective interpretation of evidence [Collective Evidence (CE), 8 items]} \\ 
\textit{Definition:} Collaborative analysis of evidence about students' experiences to reach shared understanding, guide decision-making, and implement informed changes to improve programs (\textit{e.g.,} ``Typically, systematic evidence about students' experiences in our program guides collective actions by faculty and/or staff.'')\\
\hline
\textbf{4 --- Systematic approaches to planning and monitoring change [Sustainability (S), 5 items]} \\ 
\textit{Definition:} Departmental commitment to ensuring the longevity and effectiveness of change efforts through proactive planning, systematic monitoring, assessment, and documentation (\textit{e.g.,} ``On average, people in my department build assessments into change plans.'')\\
\hline
\textbf{5 --- Fostering inclusive change to disrupt systemic injustices [Disrupting Injustices (DI), 8 items]} \\ 
\textit{Definition:} The department's commitment to centering the voices, needs, and participation of underrepresented groups in physics to address power imbalances and create a more just system (\textit{e.g.,} ``On average, people in my department take steps to ensure that marginalized people have an active role in departmental decision making.'')\\
\hline\hline

\end{tabularx}
\caption{\label{tab:definitions} Definitions of each factor alongside example items from each.}
\end{table*}

\section{\label{subsec:discussion} Discussion}
This study examined the psychometric properties of data derived from the \textit{Culture around Systemic Change Survey (CSCS),} developed to measure physics departments' approach to educational change. We evaluated the dimensional structure of the data using Exploratory Factor Analysis (EFA), estimated internal consistency reliability of factor scores and assessed convergent and divergent validity. 

Before conducting EFA, we evaluated whether the data were appropriate for factor analysis. The overall KMO measure was 0.89, indicating sampling adequacy well above the commonly accepted threshold of 0.60 and supporting the factorability of the correlation matrix. Bartlett's Test of Sphericity was also significant \textit{(p$<$0.001)}, confirming that the observed correlations between items were sufficient for factor analysis. At the item level, however, three questions (1.5, 1.13, and 1.15) demonstrated low inter-item correlations raising concerns about their contribution to shared variance. Additionally, a pair of items (1.23 and 1.24) showed signs of multicollinearity, with a very high inter-item correlation \textit{(r$=$0.88)}. As such, we excluded questions 1.5, 1.13, 1.15, and 1.23 prior to factor extraction. Moreover, although Mardia's test indicated multivariate non-normality, this results is not uncommon and can be sensitive to minor departures from normality. Visual inspection of univariate distributions showed that skewness and kurtosis were largely within acceptable ranges. In conclusion, these results support the appropriate use of EFA in the cleaned dataset.

We started EFA analysis with an evaluation of how many factor to retain. While we initially hypothesized eight constructs, we expected that conceptual overlap could yield to fewer distinct latent factors. Scree plot analysis suggested a sharp drop after the first eigenvalue, but both Horn's parallel analysis and Kaiser's criterion suggested 4 and 7 factors, respectively. Following these quantitative indicators along with initial theoretical considerations, we tested models with 1 to 8 factors. Solutions with more factors were eliminated due to insufficient item loadings per factor. 

The final 5-factor accounted for 69\% of the total variance in the retained items. Each factor was labeled based on its conceptual meaning: Open-mindedness (OM), Student Involvement (SI), Collective Interpretation of Evidence (CE), Sustainability (S), and Disruption of Systemic Injustices (DI).  All five factors exceeded the 0.50 threshold for the Average Variance Extracted (AVE) suggesting that the data support convergent validity. Moreover, the squared inter-construct correlations were lower than their corresponding AVEs indicating support for discriminant validity. Finally, internal consistency for the total score was high ($\alpha=0.97$, $\omega_t=0.98$), and a hierarchical omega of 0.88 suggests that most the reliable variance reflects a general factor. Subscale (factor) reliability estimates were also high ($\omega_{t \cdot sub}$ and $\alpha_{sub}$ between 0.89 and 0.95), but lower hierachical omega values ($\omega_{h \cdot sub}$ ranging from 0.18 to 0.40) indicate that much of this reliability stems from the individual differences of the general factor rather than differences in the specific subscales. Therefore, while the total score can be interpreted as a reliable indicator of a general construct (e.g., culture around systemic change), additional validation is needed before factor scores can be used as standalone measures of distinct dimensions.

Overall, the final final five-factor model closely reflects our initial theorization of cultural dimensions relevant to departmental change and offers strong preliminary evidence supporting the structural validity of the data interpretations. The total score demonstrates strong internal consistency and reflects a reliable general factor underlying the survey instrument. However, further work is needed further to establish the reliability of the individual subscale responses.

\subsection{Limitations}

Several limitations should be acknowledged. First, the relatively small sample size and limited number of responses per department contrain the generalizability of the findings. Second, while we conducted cognitive interviews (think alouds) with some participants during the survey development stage, additional interviews with staff members are needed to ensure item clarity and relevance across groups. Third, this study relied solely on exploratory factor analysis which is the first step of the psychometric evaluation process, a process that is iterative in nature. We collected additional data and will perform confirmatory factor analysis and internal consistency metrics to test the stability and generalizability of the proposed factor structure. This work will be the focus of a future publication and we invite more researchers to continue using and refining this instrument across various contexts.

\subsection{Implications for Research and Practice}

The findings from this study offer several important implications for both researchers and practitioners seeking to understand and shape departmental culture in higher education. Although this survey was developed with physics departments in mind, its underlying constructs are broadly relevant and should not be viewed as discipline-specific. Researchers in other fields are encouraged to review, adapt, and validate the instrument responses within varied disciplinary contexts, accounting for local cultural norms and priorities. The five-factor solution identified in this analysis provides a theoretical grounded and empirically supported framework for studying how departments approach systemic change. Results from this survey should be used to examine other important outcomes such as student retention or efficacy of change efforts. Additionally, the strong reliability of the general factor suggests that the total score may serve as a meaningful, high-level indicator of department culture and readiness for change, while the factor structure offers a foundation for future confirmatory analyses and continued refinement of the survey instrument. 

The survey instrument was developed with leaders of disciplinary change initiatives in mind as primary practitioners who can be informed by its findings. While the instrument may also be used to assess culture at the level of individual departments, we advise caution in doing so. Responses to the survey may reflect sensitive perceptions and experiences, and as such, data collection, storage, and dissemination of findings must follow rigorous ethical protocols to ensure the protection of participants' identities and confidentiality, give the small pool of respondents per department.

\section{\label{subsec:conclusions} Conclusions}

This study presents the first attempt to assess the psychometric evaluation of a newly developed survey designed to assess cultural dimensions relevant to systemic change in physics departments. Using exploratory factor analysis, we identified a five factor solution that closely aligns with our theoretical expectations. The scores gathered from its first pilot testing offered preliminary  evidence for structural validity. Future work will focus on confirmatory analyses, broader sampling, and continued refinement of the survey instrument. As development progresses, this survey tool has the potential to support both research on departmental/institutional change and practical efforts to foster a more equitable and systemic approach to the change process. This work sets the foundation for conducting population studies that measure the state of progress of the physics community along culture around systemic change. 

\begin{table*}[t]
\centering
\def\arraystretch{1.1}
\caption{\label{tab:surveyitems} List of original 50 items on the administered pilot survey}
\begin{tabularx}{\textwidth}{>{\hsize=.1\hsize}X
                             > {\hsize=.9\hsize}X}
\hline

\multicolumn{2}{l}{\textbf{On average, people in my department...}}\\
$\bullet$ 1.1 & engage with differing perspectives. \\
$\bullet$ 1.2 & take advantage of opportunities to learn, grow, or change. \\
$\bullet$ 1.3 & are open to revising their thinking. \\
$\bullet$ 1.4 & hold stubbornly to their own opinion. \\
$\bullet$ 1.5 & treat conversations as arguments to be won. \\
$\bullet$ 1.6 & discuss openly common biases in educational practices (hiring decisions, admission practices, writing and/or reading recommendation letters, etc.). \\
$\bullet$ 1.7 & partner with student representatives to collectively pursue change efforts. \\
$\bullet$ 1.8 & partner with student representatives in a way that allows students to meaningfully participate in decision making. \\
$\bullet$ 1.9 & communicate about the intended purpose of their change efforts.\\
$\bullet$ 1.10 & share the reasoning behind the changes made. \\
$\bullet$ 1.11 & utilize relevant experts (e.g., equity and inclusion experts, physics education researchers) when collecting and interpreting data.\\
$\bullet$ 1.12 & adapt change efforts to the available resources (within or external to the department).\\
$\bullet$ 1.13 & consider peoples’ power and influence when planning a change effort.\\
$\bullet$ 1.14 & take into account the current state of our program(s) when planning a change effort.\\
$\bullet$ 1.15 & are supported by the resources needed to make the change feasible.\\
$\bullet$ 1.16 & periodically revisit the justifications for past programmatic changes.\\
$\bullet$ 1.17 & systematically monitor change efforts to understand progress toward goals.\\
$\bullet$ 1.18 & build assessments into change plans.\\
$\bullet$ 1.19 & anticipate potential challenges with change plans.\\
$\bullet$ 1.20 & build plans for how to overcome potential challenges with change initiatives. \\
$\bullet$ 1.21 & do not collect systematic evidence about students’ experiences in our programs(s) unless it is mandated (by the institution, department, etc.).\\
$\bullet$ 1.22 & disrupt biases that they recognize in departmental processes. \\
$\bullet$ 1.23 & are transparent about methods of collecting data about students’ experiences in our program(s). \\
$\bullet$ 1.24 & are transparent about analysis and representation of data about students’ experiences in our program(s). \\
$\bullet$ 1.25 & monitor the effects of the change efforts on the experiences of marginalized people. \\
$\bullet$ 1.26 & take steps to ensure that marginalized people have an active role in departmental decision making. \\
$\bullet$ 1.27 & actively attends to the needs of marginalized people. \\
$\bullet$ 1.28 & take steps to ensure that marginalized people feel safe and comfortable voicing their opinions or concerns. \\
$\bullet$ 1.29 & address power differentials in departmental conversations. \\
$\bullet$ 1.30 & take action to build a more just system. \\
\hline
\multicolumn{2}{l}{\textbf{Typically, systematic evidence about students’ experiences in our program(s)...}}\\
$\bullet$ 2.1 & leads to informed decision making. \\
$\bullet$ 2.2 & leads to change(s) (e.g. modifying recruitment practices, curricular modifications, instructional changes, etc.). \\
$\bullet$ 2.3 & is used to improve the experiences of marginalized students. \\
$\bullet$ 2.4 & is interpreted in collaboration with student representatives. \\
$\bullet$ 2.5 & is used to identify the nature of problems in our program(s). \\
$\bullet$ 2.6 & is interpreted in a collaborative environment with multiple faculty members. \\
$\bullet$ 2.7 & leads to collective decision making among faculty, and/or staff.\\
$\bullet$ 2.8 & guides collective actions by faculty and/or staff. \\
\hline
\multicolumn{2}{l}{\textbf{On average, my department’s change efforts...}}\\
$\bullet$ 3.1 & are supported by formal evidence (such as outcomes from surveys, interviews, literature, expert feedback, assessments, national or institutional reports). \\
$\bullet$ 3.2 & are based on adapting research-based, well-established practices to local situations.\\
$\bullet$ 3.3 & are informed by our students’ values or goals. \\
$\bullet$ 3.4 & are based on consensus among faculty.\\
$\bullet$ 3.5 & are driven by a shared responsibility among department faculty for the health of the department and the people in it. \\
$\bullet$ 3.6 & involve students in decision making.\\
$\bullet$ 3.7 & involve students in implementing changes. \\
$\bullet$ 3.8 & are guided by informal student feedback (e.g., ad-hoc conversations).\\
$\bullet$ 3.9 & are documented for others to consider or review. \\
$\bullet$ 3.10 & are designed to rectify past injustices (e.g. discrimination, exclusion). \\
$\bullet$ 3.11 & result in changes that positively impact marginalized people. \\
$\bullet$ 3.12 & are successful in making the department a better place for its students. \\

\hline
\end{tabularx}
\end{table*}

\begin{acknowledgments}

This work is supported by the American Physical Society Innovation Fund, APS IF-13. The authors acknowledge the contributions and participation in the co-design sessions of the following people: Susan White, Patrick Mulvey, Rachel Ivie, Geraldine Cochran, Jesus Pando, Joel Corbo, David Craig, Jovonni Spinner, Mario Borunda, Arlene Knowles, Dessie Clark, Laura McCullough, Robert P. Dalka, Patrick Banner, and Tom Rice. We are also grateful to the expert panel review committee members, Courtney Ngai, Kerrie Douglas, Susan White, Rachel Ivie, and Patrick Mulvey, and external consultants, Sarah Wise and AnneMarie Vaccaro for the feedback and consultation on the item development. Finally, we thank the undergraduate student intern, Mujtaba Khalid for the valuable contribution to the development of the pilot contact list.

\end{acknowledgments}

\begin{table}[h]
\centering
\def\arraystretch{1.3}
\caption{\label{tab:correlations} Correlation matrix between all the items used in the final EFA, which allows for a reproduction of all analyses.}
\resizebox{\textwidth}{!}{%
\begin{tabular}{rrrrrrrrrrrrrrrrrrrrrrrrrrrrrrrrrrrr}
  \hline
 & \textbf{1.1} & \textbf{1.2} & \textbf{1.3} & \textbf{1.4} & \textbf{1.7} & \textbf{1.8} & \textbf{1.10} & \textbf{1.14} & \textbf{1.17} & \textbf{1.18} & \textbf{1.20} & \textbf{1.22} & \textbf{1.25} & \textbf{1.26} & \textbf{1.27} & \textbf{1.28} & \textbf{1.29} & \textbf{1.30} & \textbf{2.1} & \textbf{2.2} & \textbf{2.3} & \textbf{2.4} & \textbf{2.5} & \textbf{2.6} & \textbf{2.7} & \textbf{2.8} & \textbf{3.1} & \textbf{3.2} & \textbf{3.3} & \textbf{3.4} & \textbf{3.5} & \textbf{3.6} & \textbf{3.7} & \textbf{3.9} & \textbf{3.10} \\ 
  \hline
\textbf{1.1} & 1.00 &  &  &  &  &  &  &  &  &  &  &  &  &  &  &  &  &  &  &  &  &  &  &  &  &  &  &  &  &  &  &  &  &  &  \\ 
  \textbf{1.2} & 0.67 & 1.00 &  &  &  &  &  &  &  &  &  &  &  &  &  &  &  &  &  &  &  &  &  &  &  &  &  &  &  &  &  &  &  &  &  \\ 
  \textbf{1.3} & 0.65 & 0.80 & 1.00 &  &  &  &  &  &  &  &  &  &  &  &  &  &  &  &  &  &  &  &  &  &  &  &  &  &  &  &  &  &  &  &  \\ 
  \textbf{1.4} & 0.47 & 0.57 & 0.72 & 1.00 &  &  &  &  &  &  &  &  &  &  &  &  &  &  &  &  &  &  &  &  &  &  &  &  &  &  &  &  &  &  &  \\ 
  \textbf{1.7} & 0.43 & 0.56 & 0.41 & 0.35 & 1.00 &  &  &  &  &  &  &  &  &  &  &  &  &  &  &  &  &  &  &  &  &  &  &  &  &  &  &  &  &  &  \\ 
  \textbf{1.8} & 0.32 & 0.44 & 0.31 & 0.33 & 0.76 & 1.00 &  &  &  &  &  &  &  &  &  &  &  &  &  &  &  &  &  &  &  &  &  &  &  &  &  &  &  &  &  \\ 
  \textbf{1.10} & 0.50 & 0.53 & 0.53 & 0.41 & 0.42 & 0.29 & 1.00 &  &  &  &  &  &  &  &  &  &  &  &  &  &  &  &  &  &  &  &  &  &  &  &  &  &  &  &  \\ 
  \textbf{1.14} & 0.57 & 0.66 & 0.55 & 0.40 & 0.52 & 0.38 & 0.53 & 1.00 &  &  &  &  &  &  &  &  &  &  &  &  &  &  &  &  &  &  &  &  &  &  &  &  &  &  &  \\ 
  \textbf{1.17} & 0.36 & 0.48 & 0.39 & 0.25 & 0.51 & 0.38 & 0.47 & 0.48 & 1.00 &  &  &  &  &  &  &  &  &  &  &  &  &  &  &  &  &  &  &  &  &  &  &  &  &  &  \\ 
  \textbf{1.18} & 0.26 & 0.38 & 0.36 & 0.17 & 0.44 & 0.41 & 0.39 & 0.39 & 0.76 & 1.00 &  &  &  &  &  &  &  &  &  &  &  &  &  &  &  &  &  &  &  &  &  &  &  &  &  \\ 
  \textbf{1.20} & 0.35 & 0.58 & 0.44 & 0.22 & 0.59 & 0.42 & 0.44 & 0.53 & 0.73 & 0.72 & 1.00 &  &  &  &  &  &  &  &  &  &  &  &  &  &  &  &  &  &  &  &  &  &  &  &  \\ 
  \textbf{1.22} & 0.40 & 0.50 & 0.46 & 0.41 & 0.54 & 0.50 & 0.42 & 0.54 & 0.50 & 0.42 & 0.47 & 1.00 &  &  &  &  &  &  &  &  &  &  &  &  &  &  &  &  &  &  &  &  &  &  &  \\ 
  \textbf{1.25} & 0.33 & 0.44 & 0.39 & 0.29 & 0.60 & 0.51 & 0.34 & 0.51 & 0.59 & 0.64 & 0.65 & 0.56 & 1.00 &  &  &  &  &  &  &  &  &  &  &  &  &  &  &  &  &  &  &  &  &  &  \\ 
  \textbf{1.26} & 0.28 & 0.41 & 0.42 & 0.41 & 0.57 & 0.55 & 0.35 & 0.37 & 0.50 & 0.52 & 0.47 & 0.58 & 0.69 & 1.00 &  &  &  &  &  &  &  &  &  &  &  &  &  &  &  &  &  &  &  &  &  \\ 
  \textbf{1.27} & 0.37 & 0.45 & 0.42 & 0.32 & 0.60 & 0.56 & 0.30 & 0.44 & 0.54 & 0.56 & 0.58 & 0.63 & 0.77 & 0.77 & 1.00 &  &  &  &  &  &  &  &  &  &  &  &  &  &  &  &  &  &  &  &  \\ 
  \textbf{1.28} & 0.30 & 0.41 & 0.47 & 0.38 & 0.52 & 0.45 & 0.37 & 0.45 & 0.49 & 0.43 & 0.46 & 0.66 & 0.71 & 0.76 & 0.80 & 1.00 &  &  &  &  &  &  &  &  &  &  &  &  &  &  &  &  &  &  &  \\ 
  \textbf{1.29} & 0.32 & 0.39 & 0.37 & 0.34 & 0.50 & 0.56 & 0.31 & 0.45 & 0.40 & 0.41 & 0.52 & 0.60 & 0.62 & 0.66 & 0.69 & 0.68 & 1.00 &  &  &  &  &  &  &  &  &  &  &  &  &  &  &  &  &  &  \\ 
  \textbf{1.30} & 0.40 & 0.49 & 0.50 & 0.44 & 0.58 & 0.55 & 0.42 & 0.56 & 0.46 & 0.38 & 0.46 & 0.64 & 0.63 & 0.62 & 0.66 & 0.67 & 0.67 & 1.00 &  &  &  &  &  &  &  &  &  &  &  &  &  &  &  &  &  \\ 
  \textbf{2.1} & 0.32 & 0.53 & 0.49 & 0.34 & 0.60 & 0.53 & 0.45 & 0.52 & 0.50 & 0.53 & 0.60 & 0.46 & 0.56 & 0.45 & 0.51 & 0.54 & 0.48 & 0.54 & 1.00 &  &  &  &  &  &  &  &  &  &  &  &  &  &  &  &  \\ 
  \textbf{2.2} & 0.35 & 0.49 & 0.35 & 0.27 & 0.55 & 0.50 & 0.39 & 0.56 & 0.54 & 0.59 & 0.66 & 0.38 & 0.54 & 0.40 & 0.48 & 0.40 & 0.43 & 0.46 & 0.78 & 1.00 &  &  &  &  &  &  &  &  &  &  &  &  &  &  &  \\ 
  \textbf{2.3} & 0.34 & 0.51 & 0.44 & 0.32 & 0.63 & 0.56 & 0.37 & 0.48 & 0.50 & 0.60 & 0.60 & 0.47 & 0.72 & 0.66 & 0.70 & 0.64 & 0.61 & 0.63 & 0.76 & 0.73 & 1.00 &  &  &  &  &  &  &  &  &  &  &  &  &  &  \\ 
  \textbf{2.4} & 0.25 & 0.41 & 0.34 & 0.23 & 0.63 & 0.60 & 0.34 & 0.41 & 0.45 & 0.42 & 0.52 & 0.46 & 0.54 & 0.45 & 0.51 & 0.52 & 0.58 & 0.58 & 0.62 & 0.51 & 0.66 & 1.00 &  &  &  &  &  &  &  &  &  &  &  &  &  \\ 
  \textbf{2.5} & 0.35 & 0.49 & 0.38 & 0.31 & 0.61 & 0.50 & 0.43 & 0.55 & 0.55 & 0.51 & 0.63 & 0.48 & 0.55 & 0.50 & 0.52 & 0.58 & 0.49 & 0.56 & 0.71 & 0.73 & 0.67 & 0.58 & 1.00 &  &  &  &  &  &  &  &  &  &  &  &  \\ 
  \textbf{2.6} & 0.30 & 0.45 & 0.38 & 0.32 & 0.53 & 0.48 & 0.36 & 0.57 & 0.56 & 0.53 & 0.63 & 0.56 & 0.60 & 0.58 & 0.59 & 0.63 & 0.53 & 0.52 & 0.66 & 0.65 & 0.70 & 0.50 & 0.80 & 1.00 &  &  &  &  &  &  &  &  &  &  &  \\ 
  \textbf{2.7} & 0.35 & 0.45 & 0.35 & 0.27 & 0.53 & 0.49 & 0.34 & 0.60 & 0.50 & 0.57 & 0.63 & 0.46 & 0.63 & 0.53 & 0.57 & 0.57 & 0.59 & 0.46 & 0.70 & 0.74 & 0.76 & 0.53 & 0.71 & 0.82 & 1.00 &  &  &  &  &  &  &  &  &  &  \\ 
  \textbf{2.8} & 0.29 & 0.41 & 0.32 & 0.22 & 0.53 & 0.51 & 0.28 & 0.51 & 0.55 & 0.60 & 0.64 & 0.46 & 0.62 & 0.52 & 0.59 & 0.61 & 0.57 & 0.49 & 0.70 & 0.72 & 0.75 & 0.55 & 0.76 & 0.83 & 0.92 & 1.00 &  &  &  &  &  &  &  &  &  \\ 
  \textbf{3.1} & 0.41 & 0.38 & 0.28 & 0.16 & 0.48 & 0.39 & 0.27 & 0.37 & 0.55 & 0.61 & 0.61 & 0.37 & 0.48 & 0.40 & 0.43 & 0.39 & 0.41 & 0.44 & 0.61 & 0.63 & 0.60 & 0.43 & 0.67 & 0.53 & 0.57 & 0.60 & 1.00 &  &  &  &  &  &  &  &  \\ 
  \textbf{3.2} & 0.53 & 0.56 & 0.43 & 0.34 & 0.67 & 0.58 & 0.31 & 0.60 & 0.51 & 0.50 & 0.61 & 0.57 & 0.62 & 0.47 & 0.60 & 0.49 & 0.56 & 0.55 & 0.63 & 0.65 & 0.63 & 0.62 & 0.60 & 0.59 & 0.62 & 0.58 & 0.61 & 1.00 &  &  &  &  &  &  &  \\ 
  \textbf{3.3} & 0.51 & 0.67 & 0.52 & 0.37 & 0.66 & 0.58 & 0.47 & 0.63 & 0.58 & 0.55 & 0.61 & 0.61 & 0.56 & 0.45 & 0.52 & 0.51 & 0.54 & 0.61 & 0.65 & 0.66 & 0.68 & 0.60 & 0.67 & 0.58 & 0.60 & 0.62 & 0.61 & 0.72 & 1.00 &  &  &  &  &  &  \\ 
  \textbf{3.4} & 0.38 & 0.45 & 0.53 & 0.34 & 0.32 & 0.29 & 0.31 & 0.38 & 0.38 & 0.39 & 0.43 & 0.35 & 0.35 & 0.36 & 0.37 & 0.45 & 0.31 & 0.31 & 0.52 & 0.41 & 0.49 & 0.31 & 0.46 & 0.63 & 0.59 & 0.60 & 0.41 & 0.45 & 0.34 & 1.00 &  &  &  &  &  \\ 
  \textbf{3.5} & 0.49 & 0.59 & 0.61 & 0.44 & 0.46 & 0.39 & 0.32 & 0.56 & 0.51 & 0.49 & 0.57 & 0.56 & 0.51 & 0.51 & 0.55 & 0.53 & 0.47 & 0.52 & 0.62 & 0.51 & 0.61 & 0.39 & 0.54 & 0.64 & 0.62 & 0.62 & 0.56 & 0.65 & 0.56 & 0.66 & 1.00 &  &  &  &  \\ 
  \textbf{3.6} & 0.35 & 0.55 & 0.40 & 0.36 & 0.72 & 0.61 & 0.24 & 0.46 & 0.39 & 0.36 & 0.56 & 0.44 & 0.49 & 0.49 & 0.56 & 0.45 & 0.47 & 0.52 & 0.58 & 0.49 & 0.59 & 0.63 & 0.54 & 0.54 & 0.53 & 0.54 & 0.46 & 0.64 & 0.59 & 0.32 & 0.58 & 1.00 &  &  &  \\ 
  \textbf{3.7} & 0.46 & 0.55 & 0.39 & 0.36 & 0.67 & 0.56 & 0.30 & 0.49 & 0.40 & 0.41 & 0.54 & 0.44 & 0.47 & 0.40 & 0.47 & 0.39 & 0.47 & 0.43 & 0.56 & 0.53 & 0.55 & 0.61 & 0.55 & 0.52 & 0.56 & 0.52 & 0.52 & 0.68 & 0.62 & 0.34 & 0.53 & 0.77 & 1.00 &  &  \\ 
  \textbf{3.9} & 0.27 & 0.31 & 0.25 & 0.22 & 0.50 & 0.49 & 0.29 & 0.38 & 0.56 & 0.58 & 0.51 & 0.38 & 0.52 & 0.47 & 0.50 & 0.37 & 0.43 & 0.46 & 0.49 & 0.53 & 0.51 & 0.48 & 0.64 & 0.57 & 0.55 & 0.62 & 0.57 & 0.51 & 0.51 & 0.31 & 0.34 & 0.43 & 0.48 & 1.00 &  \\ 
  \textbf{3.10} & 0.20 & 0.29 & 0.26 & 0.08 & 0.55 & 0.46 & 0.25 & 0.35 & 0.40 & 0.38 & 0.46 & 0.49 & 0.56 & 0.53 & 0.63 & 0.54 & 0.53 & 0.54 & 0.43 & 0.32 & 0.54 & 0.48 & 0.49 & 0.47 & 0.46 & 0.48 & 0.43 & 0.48 & 0.45 & 0.19 & 0.32 & 0.50 & 0.38 & 0.55 & 1.00 \\ 
   \hline
\end{tabular}%
}
\end{table}

\begin{table*}[t]
\renewcommand\tabularxcolumn[1]{m{#1}}
\centering
\def\arraystretch{1.1}
\caption{\label{tab:EFAcomparison} A comparison of factor loadings based on several methods of missing data handling and sample sizes.  Results remain highly consistent regardless of method used.  The Sustainability factor is most different, with one to two items being swapped from one solution to another.}
\begin{tabularx}{\textwidth}{|>{\hsize=1\hsize}Y|
                               > {\hsize=1\hsize}Y
                               > {\hsize=1\hsize}Y
                               > {\hsize=1\hsize}Y|
                               > {\hsize=1\hsize}Y
                               > {\hsize=1\hsize}Y
                               > {\hsize=1\hsize}Y|
                               > {\hsize=1\hsize}Y
                               > {\hsize=1\hsize}Y
                               > {\hsize=1\hsize}Y|
                               > {\hsize=1\hsize}Y
                               > {\hsize=1\hsize}Y
                               > {\hsize=1\hsize}Y|
                               > {\hsize=1\hsize}Y
                               > {\hsize=1\hsize}Y
                               > {\hsize=1\hsize}Y|}
\cline{1-16}
& \multicolumn{3}{c|}{\textbf{OM}} & \multicolumn{3}{c|}{\textbf{SI}} & \multicolumn{3}{c|}{\textbf{CE}} & \multicolumn{3}{c|}{\textbf{S}} & \multicolumn{3}{c|}{\textbf{DI}} \\  

\textbf{Item} & 
ML1$^{\ast}$ & ML2$^{\dagger}$ & MI$^{\ddagger}$  &
ML1  & ML2  & MI  & 
ML1  & ML2  & MI  &
ML1  & ML2  & MI  &
ML1  & ML2  & MI 
\\ \hline


\textbf{1.3} & 
\textbf{0.99} & \textbf{0.85} & \textbf{0.94} & 
-0.15 & -0.07 & -0.09 & 
-0.08 & -0.06 & -0.03 &
0.02 & 0.01 & -0.03 &
0.09 & 0.10 & 0.08 
\\ 

\textbf{1.2} & 
\textbf{0.79} & \textbf{0.66} & \textbf{0.78} & 
0.29 & 0.26 & 0.28 &
-0.09 & -0.11 & -0.07 &
0.07 & 0.21 & 0.08 &
-0.16 & -0.10 & -0.16  
\\ 

\textbf{1.4} & 
\textbf{0.77} & \textbf{0.77} & \textbf{0.71} &
0.02 & 0.04 & 0.04 &
-0.05 & 0.01 & -0.04 &
-0.25 & -0.32 & -0.22 &
0.18 & 0.15 & 0.15
\\ 

\textbf{1.1} & 
\textbf{0.69} & \textbf{0.69} & \textbf{0.59} &
0.17 & 0.21 & 0.21 &
-0.11 & -0.17 & -0.21 &
0.06 & 0.15 & 0.16 &
-0.09 & -0.17 & 0.01
\\ 

\textbf{1.10} & 
\textbf{0.50} & \textbf{0.45} & 0.39 &
-0.01 & -0.05 & -0.02 &
-0.17 & -0.10 & -0.16 &
0.32 & 0.37 & \textbf{0.51} &
0.04 & 0.05 & 0.00 
\\ 

\textbf{3.5} 
& \textbf{0.44} & \textbf{0.42} & \textbf{0.52} &
-0.02 & 0.03 & -0.04 &
0.39 & \textbf{0.44} & \textbf{0.47} &
0.03 & -0.06 & -0.15 &
0.10 & 0.11 & 0.09
\\

\textbf{1.14} & 
\textbf{0.42} & NA & \textbf{0.54} &
0.24 & NA & 0.06 &
0.20 & NA & 0.10 &
0.05 & NA & 0.21 &
-0.05 & NA & -0.04 
\\

& & & & & & & & & & & & & & & \\
\textbf{3.6} & 
0.05 & 0.04 & 0.07 &
\textbf{0.95} & \textbf{0.88} & \textbf{0.89} &
0.04 & 0.07 & 0.09 &
-0.26 & -0.19 & -0.29 &
0.00 & 0.00 & -0.01 
\\

\textbf{3.7} & 
0.11 & 0.14 & 0.16 &
\textbf{0.92} & \textbf{0.88} & \textbf{0.83} &
0.04 & 0.05 & 0.04 &
-0.11 & -0.10 & -0.16 &
-0.18 & -0.17 & -0.13 
\\

\textbf{1.7}  & 
0.02 & 0.01 & 0.04 &
\textbf{0.91} & \textbf{0.82} & \textbf{0.87} &
-0.17 & -0.10 & -0.16 &
-0.01 & -0.03 & 0.06 &
0.09 & 0.13 & 0.03 
\\

\textbf{1.8} & 
-0.10 & -0.06 & -0.09 &
\textbf{0.84} & \textbf{0.80} & \textbf{0.80} &
-0.07 & -0.05 & -0.08 &
-0.14 & -0.16 & -0.07 &
0.19 & 0.24 & 0.19
\\

\textbf{2.4} & 
-0.13 & -0.14 & -0.14 &
\textbf{0.79} & \textbf{0.67} & \textbf{0.55} &
0.04 & 0.10 & 0.13 &
-0.03 & -0.14 & 0.02 &
0.11 & 0.14 & 0.19 \\

\textbf{3.2} & 
0.14 & 0.10 & 0.18 &
\textbf{0.65} & \textbf{0.63} & \textbf{0.55} &
0.03 & 0.12 & 0.11 &
0.14 & 0.11 & 0.04 &
-0.04 & -0.06 & 0.00 
\\

\textbf{3.3} & 
0.20 & 0.15 & 0.23 &
\textbf{0.56} & \textbf{0.59} & \textbf{0.51} &
0.03 & 0.09 & 0.05 &
0.22 & 0.20 & 0.16 &
-0.06 & -0.12 & -0.06
\\

& & & & & & & & & & & & & & & \\
\textbf{2.8} & 
-0.18 & -0.18 & -0.19 &
0.02 & 0.09 & 0.02 &
\textbf{0.95} & \textbf{0.85} & \textbf{0.90} &
0.02 & 0.08 & 0.09 &
0.11 & 0.10 & 0.07 
\\ 

\textbf{2.7} & 
-0.09 & -0.10 & -0.05 &
0.04 & 0.08 & -0.01 &
\textbf{0.95} & \textbf{0.81} & \textbf{0.86} &
-0.04 & 0.10 & 0.04 &
0.07 & 0.06 & 0.08
\\

\textbf{2.6} & 
-0.02 & -0.03 & 0.03 &
0.03 & 0.01 & -0.09 &
\textbf{0.77} & \textbf{0.74} & \textbf{0.71} &
0.00 & 0.05 & 0.09 &
0.16 & 0.19 & 0.19
\\

\textbf{3.4} & 
0.39 & 0.38 & 0.24 &
-0.36 & -0.30 & -0.19 &
\textbf{0.74} & \textbf{0.68} & \textbf{0.71} &
-0.09 & -0.11 & -0.11 &
0.05 & 0.11 & 0.09 
\\

\textbf{2.5} & 
-0.04 & -0.06 & -0.02 &
0.35 & 0.37 & 0.23 &
\textbf{0.49} & \textbf{0.51} & \textbf{0.52} &
0.12 & 0.12 & 0.20 &
-0.02 & -0.07 & -0.04 
\\

\textbf{2.2} & 
-0.04 & -0.06 & -0.03 &
0.37 & 0.37 & 0.36 &
\textbf{0.47} & \textbf{0.53} & \textbf{0.50} &
0.27 & 0.27 & 0.26 &
-0.23 & -0.26 & -0.29 
\\

\textbf{2.1} & 
0.09 & 0.08 & 0.12 &
0.36 & 0.38 & 0.29 &
\textbf{0.45} & \textbf{0.52} & \textbf{0.55} &
0.05 & 0.03 & 0.01 &
-0.04 & -0.09 & -0.08 
\\

\textbf{2.3} & 
-0.05 & -0.06 & -0.09 &
0.30 & 0.26 & 0.23 &
\textbf{0.40} & \textbf{0.47} & \textbf{0.51} &
0.03 & 0.04 & -0.05 &
0.27 & 0.28 & 0.32
\\

& & & & & & & & & & & & & & & \\
\textbf{1.18} & 
-0.03 & -0.08 & -0.14 &
-0.25 & -0.16 & -0.16 &
-0.01 & 0.13 & 0.17 &
\textbf{0.98} & \textbf{0.83} & \textbf{0.82} &
0.12 & 0.11 & 0.07 
\\ 

\textbf{1.17} & 
0.10 & 0.05 & 0.06 &
-0.16 & -0.16 & -0.17 &
-0.13 & 0.01 & -0.05 &
\textbf{0.94} & \textbf{0.84} & \textbf{0.93} &
0.11 & 0.11 & 0.02 
\\

\textbf{1.20} & 
0.08 & -0.02 & 0.08 &
0.10 & 0.07 & 0.04 &
0.07 & 0.11 & 0.16 &
\textbf{0.70} & \textbf{0.81} & \textbf{0.70} &
-0.03 & -0.07 & -0.08
\\

\textbf{3.1} & 
-0.04 & NA & NA &
0.24 & NA & NA &
0.17 & NA & NA &
\textbf{0.54} & NA & NA &
-0.13 & NA & NA 
\\

\textbf{3.9} & 
-0.17 & NA & -0.17 &
0.21 & NA & 0.26 &
0.18 & NA & 0.15 &
\textbf{0.46} & NA & \textbf{0.43} &
0.06 & NA & 0.01 
\\
\textbf{1.19} & 
NA & 0.18 & 0.32 &
NA & -0.06 & -0.18 &
NA & 0.02 & 0.09 &
NA & \textbf{0.56} & \textbf{0.45} &
NA & 0.04 & 0.08 
\\

& & & & & & & & & & & & & & & \\
\textbf{1.28} & 
0.08 & 0.08 & 0.11 &
-0.18 & -0.18 & -0.16 &
0.22 & 0.21 & 0.20 &
-0.07 & -0.07 & -0.11 &
\textbf{0.90} & \textbf{0.90} & \textbf{0.91} 
\\ 

\textbf{1.26} & 
0.04 & 0.05 & 0.01 &
-0.09 & -0.09 & -0.05 &
0.02 & 0.07 & 0.03 &
0.06 & -0.03 & 0.02 &
\textbf{0.85} & \textbf{0.88} & \textbf{0.82} 
\\

\textbf{1.27} & 
-0.03 & -0.03 & -0.04 &
0.01 & 0.07 & 0.12 &
-0.02 & -0.03 & -0.01 &
0.15 & 0.17 & 0.06 &
\textbf{0.83} & \textbf{0.76} & \textbf{0.78} 
\\

\textbf{1.29} & 
-0.02 & -0.01 & -0.03 &
0.24 & 0.19 & 0.12 &
0.12 & 0.03 & 0.05 &
-0.11 & 0.01 & -0.05 &
\textbf{0.62} & \textbf{0.62} & \textbf{0.69} 
\\

\textbf{1.25} & 
-0.09 & -0.06 & -0.05 &
0.01 & 0.03 & 0.05 &
0.03 & 0.07 & 0.09 &
\textbf{0.40} & 0.36 & 0.26 &
\textbf{0.51} & \textbf{0.55} & \textbf{0.54} 
\\

\textbf{1.30} & 
0.18 & 0.18 & 0.21 &
0.33 & 0.31 & 0.26 &
-0.12 & -0.10 & -0.11 &
-0.03 & -0.03 & 0.01 &
\textbf{0.55} & \textbf{0.54} & \textbf{0.50} 
\\

\textbf{1.22} & 
0.24 & 0.23 & 0.28 &
0.19 & 0.17 & 0.07 &
-0.10 & -0.10 & -0.11 &
0.07 & 0.04 & 0.09 &
\textbf{0.50} & \textbf{0.52} & \textbf{0.48} 
\\

\textbf{3.10} & 
-0.16 & -0.20 & -0.22 &
0.34 & 0.38 & 0.26 &
-0.02 & -0.10 & -0.06 &
0.08 & 0.14 & 0.16 &
\textbf{0.47} & \textbf{0.46} & \textbf{0.43} 
\\

\hline
\multicolumn{16}{L{\textwidth}}{
\footnotesize{$^{\ast}$ ML1: Same as in Results Section \ref{subsec:results}, provided here for reference.  Uses the \textit{umx} package and a sample of $N=111$.  This method drops items that will not be used in EFA before using FIML to calculate a correlation matrix.  EFA is carried out in the \textit{umx} package as well.}}

\\
\multicolumn{16}{L{\textwidth}}{
\footnotesize{$^{\dagger}$ ML2: Uses the \textit{psych} package and a sample of $N=124$. Calculates a FIML correlation matrix based on all items in the data, then items are dropped from the correlation matrix.  EFA implemented with the \textit{psych} package as well.}}

\\
\multicolumn{16}{L{\textwidth}}{
\footnotesize{$^{\ddagger}$ MI: Uses the \textit{mice} package and a sample of $N=124$. Missing data is imputed in 20 datasets. Predictors included all other item responses and demographic variables. Pooled covariance matrices were calculated and used as input for EFA implemented in the \textit{psych} package.}}

\end{tabularx}
\end{table*}

\begin{table*}[t]
\renewcommand\tabularxcolumn[1]{m{#1}}
\centering
\def\arraystretch{1.1}
\caption{\label{tab:schmid} Factor loadings following Schmid-Leiman Transformation.  Loadings under 0.20 are suppressed for readability. \cite{schmid1957development}}
\begin{tabularx}{\columnwidth}{>{\hsize=1\hsize}X
                               > {\hsize=1\hsize}Y
                               > {\hsize=1\hsize}Y
                               > {\hsize=1\hsize}Y
                               > {\hsize=1\hsize}Y
                               > {\hsize=1\hsize}Y
                               > {\hsize=1\hsize}Y
                               > {\hsize=1\hsize}Y
                               > {\hsize=1\hsize}Y}
\cline{1-9}
\textbf{Item} & \textbf{Gen.} & \textbf{OM} & \textbf{SI} & \textbf{CE} & \textbf{S} & \textbf{DI} & \textbf{h2} & \textbf{p2} \\ \hline 

\textbf{1.3} & 0.54 & 0.75 &  &  &  &  & 0.86 & 0.33 \\ 
\textbf{1.2} & 0.65 & 0.60 &  &  &  & & 0.81 & 0.52\\ 
\textbf{1.4} & 0.40 & 0.58 &  &  &  & & 0.55 & 0.30\\ 
\textbf{1.1} & 0.50 & 0.52 &  &  &  & & 0.53 & 0.47\\ 
\textbf{1.10} & 0.48 & 0.38 &  &  &  & & 0.40 & 0.57\\ 
\textbf{3.5} & 0.69 & 0.33 &  & 0.22 &  & & 0.65 &0.75 
\\
\textbf{1.14} & 0.65 & 0.32 &  &  &  & & 0.55 & 0.78\\


\textbf{3.6} & 0.70 &  & 0.42 &  &  & &0.70 &0.71\\ 
\textbf{3.7} & 0.70 &  & 0.41 &  &  & &0.68&0.73\\
\textbf{1.7}  & 0.74 &  & 0.40 &  &  & &0.75&0.74\\
\textbf{1.8} & 0.66 &  & 0.38 &  &  & &0.61&0.70\\
\textbf{2.4} & 0.70 &  & 0.35 &  &  & &0.63&0.78\\
\textbf{3.2} & 0.79 &  & 0.29 &  &  & &0.73&0.86\\
\textbf{3.3} & 0.80 &  & 0.25 &  &  & &0.74&0.86\\


\textbf{2.8} & 0.79 &  &  & 0.52 &  & &0.91&0.68\\ 
\textbf{2.7} & 0.78 &  &  & 0.52 &  & &0.89&0.69\\
\textbf{2.6} & 0.77 &  &  & 0.42 &  & &0.78&0.76\\
\textbf{3.4} & 0.51 &  &  & 0.41 &  & &0.55&0.48\\
\textbf{2.5} & 0.78 &  &  & 0.27 &  & &0.71&0.86\\
\textbf{2.2} & 0.75 &  &  & 0.26 &  & &0.72&0.79\\
\textbf{2.1} & 0.76 &  &  & 0.25 &  & &0.67&0.87\\
\textbf{2.3} & 0.80 &  &  & 0.22 &  & &0.73&0.86\\


\textbf{1.18} & 0.67 &  &  &  & 0.57 & &0.79&0.58\\ 
\textbf{1.17} & 0.67 &  &  &  & 0.52 & &0.73&0.62\\
\textbf{1.20} & 0.76 &  &  &  & 0.39 & &0.73&0.78\\
\textbf{3.1} & 0.69 &  &  &  & 0.30 & &0.60&0.78\\
\textbf{3.9} & 0.65 &  &  &  & 0.26 & &0.53&0.79\\


\textbf{1.28} & 0.66 &  &  &  &  & 0.62 &0.84& 0.52\\ 
\textbf{1.26} & 0.63 &  &  &  &  & 0.59 &0.74& 0.53\\
\textbf{1.27} & 0.70 &  &  &  &  & 0.58 &0.83& 0.60\\
\textbf{1.29} & 0.66 &  &  &  &  & 0.43 &0.64& 0.68\\
\textbf{1.25} & 0.73 &  &  &  & 0.22 & 0.41 &0.75& 0.72\\
\textbf{1.30} & 0.69 &  &  &  &  & 0.38 &0.68& 0.69\\
\textbf{1.22} & 0.65 &  &  &  &  & 0.34 &0.59& 0.71\\
\textbf{3.10} & 0.59 &  &  &  &  & 0.33 &0.50& 0.70\\
\hline
\end{tabularx}
\end{table*}

\begin{table*}[t]
\renewcommand\tabularxcolumn[1]{m{#1}}
\centering
\def\arraystretch{1.1}
\caption{\label{tab:indcluster} Factor loadings obtained using an independent group solution, which forces the cross-loadings on factors to zero.}
\begin{tabularx}{\columnwidth}{>{\hsize=1\hsize}X
                               > {\hsize=1\hsize}Y
                               > {\hsize=1\hsize}Y
                               > {\hsize=1\hsize}Y
                               > {\hsize=1\hsize}Y
                               > {\hsize=1\hsize}Y}
\cline{1-6}
\textbf{Item} & \textbf{OM} & \textbf{SI} & \textbf{CE} & \textbf{S} & \textbf{DI} \\ \hline 

\textbf{1.3}  & 0.87 &  &  &  &   \\ 
\textbf{1.2}  & 0.90 &  &  &  & \\ 
\textbf{1.4}  & 0.68 &  &  &  & \\ 
\textbf{1.1}  & 0.74 &  &  &  & \\ 
\textbf{1.10}  & 0.61 &  &  &  & \\ 
\textbf{3.5}  & 0.71 &  &  &  &  \\
\textbf{1.14}  & 0.73 &  &  &  & \\


\textbf{3.6}  &  & 0.82 &  &  & \\ 
\textbf{3.7}  &  & 0.81 &  &  & \\
\textbf{1.7}   &  & 0.86 &  &  & \\
\textbf{1.8}  &  & 0.76 &  &  & \\
\textbf{2.4}  &  & 0.76 &  &  & \\
\textbf{3.2}  &  & 0.84 &  &  & \\
\textbf{3.3}  &  & 0.81 &  &  & \\


\textbf{2.8}  &  &  & 0.93 &  & \\ 
\textbf{2.7}  &  &  & 0.93 &  & \\
\textbf{2.6}  &  &  & 0.89 &  & \\
\textbf{3.4}  &  &  & 0.61 &  & \\
\textbf{2.5}  &  &  & 0.83 &  & \\
\textbf{2.2}  &  &  & 0.82 &  & \\
\textbf{2.1}  &  &  & 0.80 &  & \\
\textbf{2.3}  &  &  & 0.84 &  & \\


\textbf{1.18}  &  &  &  & 0.84 & \\ 
\textbf{1.17}  &  &  &  & 0.83 & \\
\textbf{1.20}  &  &  &  & 0.86 & \\
\textbf{3.1}  &  &  &  & 0.76 & \\
\textbf{3.9}  &  &  &  & 0.72 & \\


\textbf{1.28}  &  &  &  &  & 0.88 \\ 
\textbf{1.26}  &  &  &  &  & 0.83 \\
\textbf{1.27}  &  &  &  &  & 0.91 \\
\textbf{1.29}  &  &  &  &  & 0.79 \\
\textbf{1.25}  &  &  &  &  & 0.83 \\
\textbf{1.30}  &  &  &  &  & 0.77 \\
\textbf{1.22}  &  &  &  &  & 0.73 \\
\textbf{3.10}  &  &  &  &  & 0.69 \\
\hline
\end{tabularx}
\end{table*}

\begin{figure*}[h]
\centering
\includegraphics[]{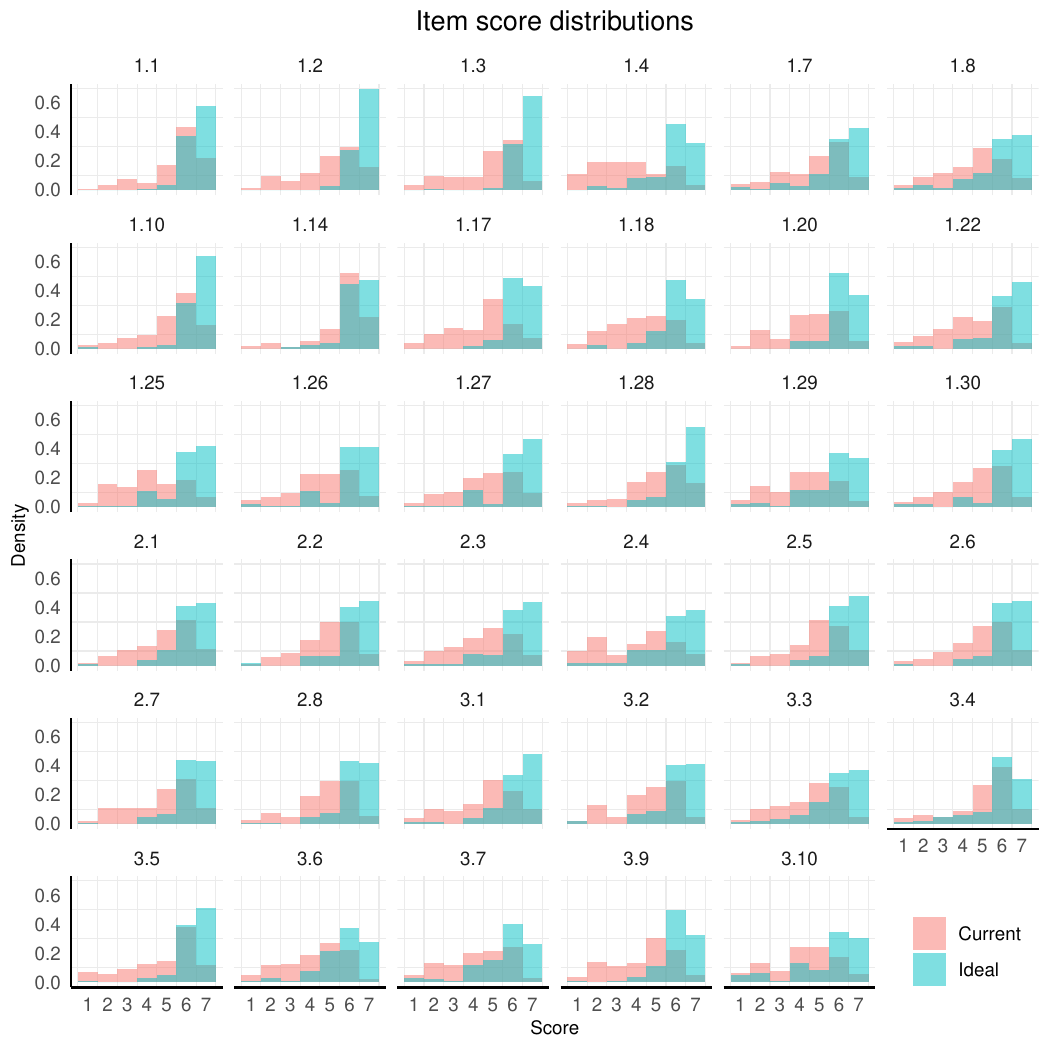}
\caption{\label{fig:distributions} Individual item score distributions across the current and ideal scales.  The ideal items are skewed toward the high end of the score distribution, which precluded their use in a factor analysis.  Future work will explore the best use of these scores.}
\end{figure*}

\bibliography{apssamp}

\end{document}